\documentclass[11pt, letterpaper, reprint, nofootinbib, notitlepage, superscriptaddress, floatfix, aps, prl]{revtex4-2}

\usepackage{amsmath, amssymb, amsfonts}
\usepackage{empheq}
\usepackage{physics}
\usepackage{amsthm}

\usepackage{subdepth}
\usepackage{bm}
\renewcommand{\mathbf}{\bm}
\usepackage{dsfont}
\renewcommand{\mathbb}{\mathds}
\usepackage[svgnames, dvipsnames]{xcolor}
\usepackage{graphicx}
\definecolor{NewBlue}{rgb}{0.1, 0.1, 0.7}
\definecolor{NewRed}{rgb}{0.7, 0.1, 0.1}
\usepackage[colorlinks, linkcolor=Maroon, citecolor=NewBlue, urlcolor=NewRed]{hyperref}
\usepackage{cleveref}

\renewcommand{\t}[1]{\mathrm{#1}}

\newcommand{\LigoMIT}{LIGO Laboratory, Massachusetts Institute of Technology, Cambridge, MA 02139}
\newcommand{\MechMIT}{Department of Mechanical Engineering, Massachusetts Institute of Technology, Cambridge, MA 02139}
\newcommand{\EqualContrib}{These authors contributed equally to this work}
\makeatletter
\newcommand{\EqualContribMark}{\textsuperscript{\normalfont\@fnsymbol{1}}}
\makeatother

\begin{document}
    % ---------- Title & authors ----------
    \title{A Quantum-Enhanced Feedback Oscillator}
    \author{Hudson A. Loughlin}
    \thanks{\EqualContrib}
    \email{hloughlin@mit.edu} \affiliation{\LigoMIT}
    \author{Daniel M. DeSantis\EqualContribMark}
    \affiliation{\LigoMIT}
    \author{Nergis Mavalvala}
    \affiliation{\LigoMIT}
    \author{Vivishek Sudhir}
    \affiliation{\LigoMIT} \affiliation{\MechMIT}

    \date{\today}

    % ---------- Abstract ----------
    \begin{abstract}
        Feedback oscillators, such as lasers and masers, serve as time references in modern computing,
        communication, and measurement. Quantum fluctuations ultimately limit their phase stability and ability to keep time precisely; 
        in the absence of quantum engineering, their phase stability is bounded by a standard quantum limit (SQL).
        Techniques to improve the frequency
        stability of feedback oscillators beyond the SQL have been theorized, but have not yet been demonstrated. We
        demonstrate an opto-electronic oscillator (OEO), a type of feedback oscillator, with phase stability
        approaching the SQL. We then engineer the OEO's quantum state to improve its phase stability, thereby
        demonstrating the essential principle quantum-enhancement of feedback oscillators. Similar techniques may be employed
        to evade the SQL in other feedback oscillators such as masers and lasers.
    \end{abstract}

    \maketitle

    % ---------- Main text ----------
    \textit{Introduction. } The generation, dissemination, and synchronization of timing signals enables much of modern
    science and technology, including fundamental measurements \cite{Ludlow15,Wiens16,Reid15,Abich19,WilkHolz12,Clivati17,Cahillane21,CalDesch22,Bruning22},
    and the processing \cite{Lind72,Xan09,Murm15,wineland98,Ball16} and communication \cite{Vit66,MeyAsch90,Kroup03} of
    information. The precision of timing signals traces back to the phase stability of the oscillators that generate them.
    Advances in timing have so far come from the development of new types of oscillators
    \cite{huygens1673,cady1922piezoelectric,Varian1939,Boot1946,Essen1955,Goldenberg1960,Ludlow15} and their isolation
    from environmental disturbances \cite{armstrong1936,Udem2002,Robinson19}. However, modern oscillators have begun to
    brush up against fundamental quantum fluctuations \cite{Sieg68,Fritschel1989,Benmessai2008,Breeze18}, present even when
    they are perfectly isolated from their environment.

    Most oscillators, including mechanical \cite{huygens1673,Rawl48}, electronic \cite{HajLee03}, electro-mechanical
    \cite{cady1922piezoelectric},
    optical \cite{SchaTow58,Maiman1960,Slusher1999}, or opto-electronic (OEOs) \cite{YaoMal96b,Maleki11}, are examples of
    feedback oscillators. A feedback oscillator essentially consists of a frequency selective element that sets its output
    frequency, an out-coupler that enables extraction of a timing signal from it, and an amplifier that
    keeps its internal energy constant. Unavoidable quantum fluctuations in the amplifier and out-coupler fundamentally
    limit the phase stability of any feedback oscillator \cite{Loughlin23,Loughlin26}. But this limit is malleable
    if the quantum state of the oscillator is engineered appropriately. Thus, quantum-enhancement offers a promising
    route to improving the performance of a wide class of oscillators, as long as they are limited by quantum fluctuations.

    We detail what is, to our knowledge, the first quantum-enhanced feedback oscillator in the form of an OEO with quantum-enhanced
    phase stability. Since their conception, OEOs have stood out among feedback oscillators for the degree to which the sources
    of their phase instability can be understood and accurately modeled \cite{YaoMal96a,YaoMal96b,Romisch1999,Romisch2000}.
    Indeed, we find that our OEO's phase noise agrees well with an ab-initio model with no free parameters. As
    anticipated by our noise model, our OEO's phase stability is limited by quantum fluctuations. Injecting squeezed
    light allows us to engineer these fluctuations and realize a feedback oscillator with quantum-enhanced phase
    stability. In doing so, we demonstrate that it is possible to enhance the phase stability of a feedback
    oscillator through quantum engineering. Similar techniques applied to lasers may allow them to achieve performance beyond
    the Schawlow-Townes limit \cite{SchaTow58,Loughlin23}.

    \begin{figure}[t]
        \centering
        \includegraphics[width=\columnwidth]{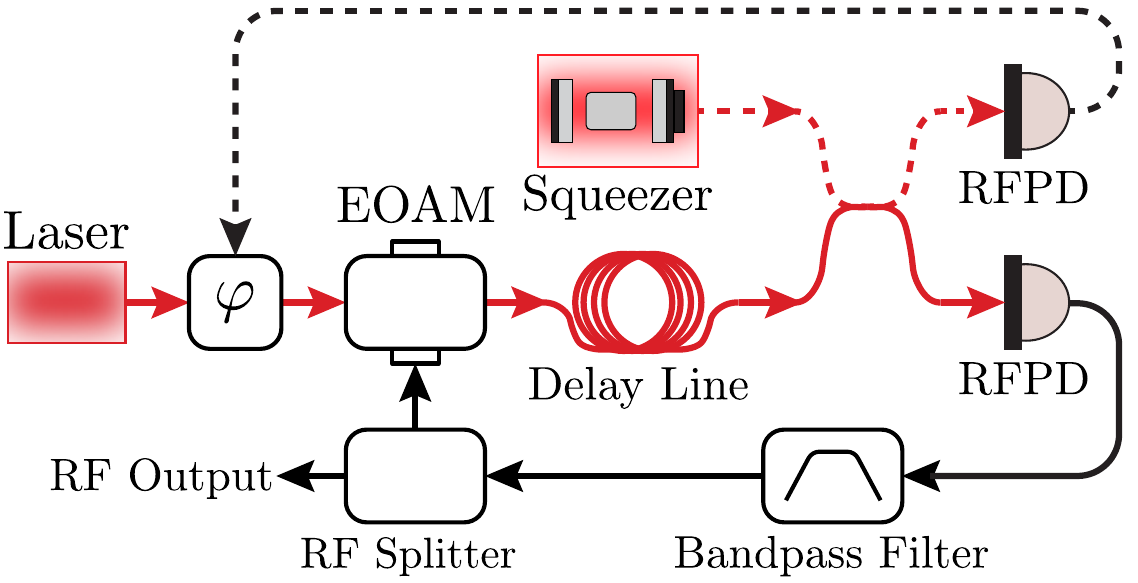}
        \caption{Experimental schematic of the quantum-enhanced opto-electronic oscillator. 
        Red lines denote optical signals and black lines denote electronic signals. 
        Dashed lines denote signal paths that are only connected when squeezed light is injected into the OEO.}
        \label{fig:simple_schematic}
    \end{figure}

    \begin{figure*}[t]
        \centering
        \includegraphics[width=\textwidth]{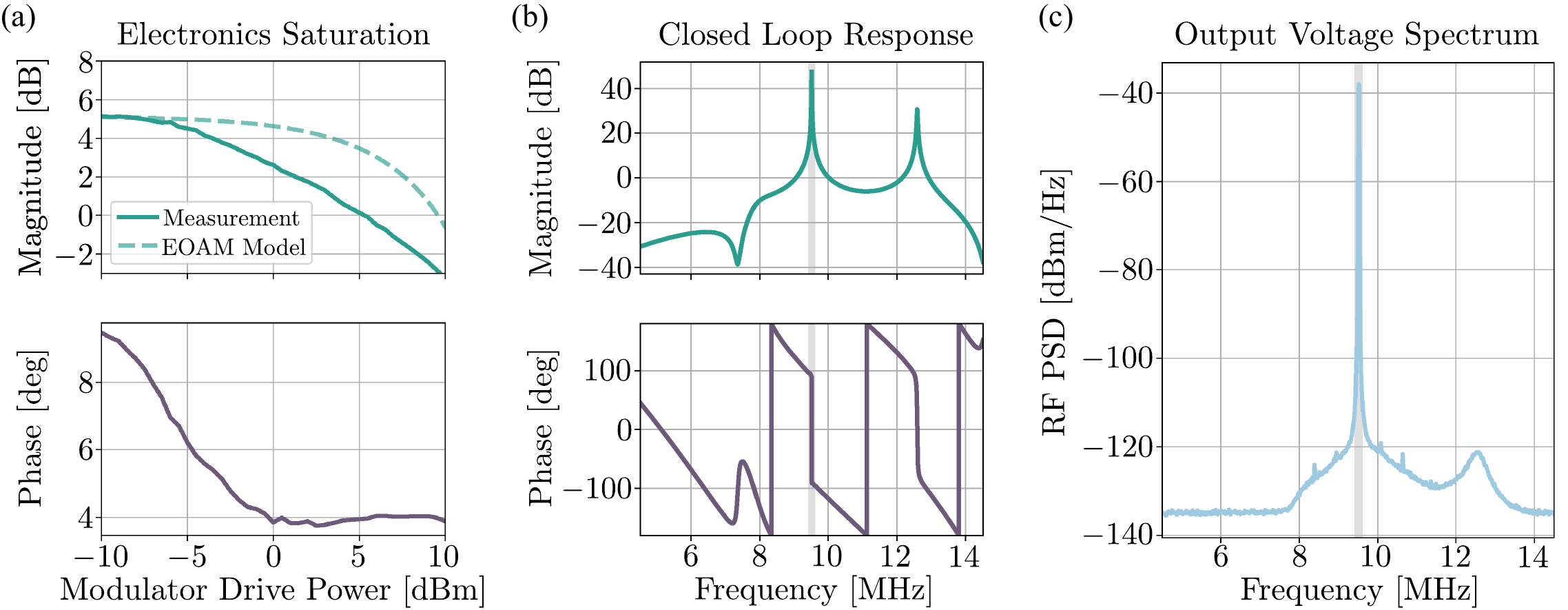}
        \caption{OEO characterization. (a) Saturation effects in the OEO's open loop transfer function at a frequency of 9.5 MHz and with 65 $\mu$W of average optical power on the in-loop detector. 
        Top shows the RF to optical gain as the RF drive power is increased: 
        blue dashed in a model of saturation from the EOAM's sinusoidal response, blue solid is the
        measured gain, likely dominated by the nonlinearity of the photodetector's transimpedance
        response. The measurement predicts the OEO will output about 5 dBm of RF power. 
        Bottom shows the phase response as the RF drive power is increased, which we attribute to the
        photodetector's finite slew rate.
        (b) Closed loop OEO response inferred from open loop measurements. These measurements predict the OEO will oscillate at 9.51 MHz. (c) shows the OEO's output voltage spectrum, peaking 
        at 9.52 MHz with an integrated output power of 6.6 dBm between 9.41 and 9.61 MHz. The grey bands in (b) and (c) indicate the region within 100 kHz of the resonance frequency.}
        \label{fig:classical_characterization}
    \end{figure*}

    \textit{Principle of the OEO.} An OEO is an electro-optic feedback loop configured as an oscillator
    \cite{YaoMal96a,YaoMal96b,Romisch1999,Romisch2000}. \Cref{fig:simple_schematic} shows 
    the schematic of such an oscillator: a laser beam is sent through an electro-optic amplitude 
    modulator (EOAM) and a fiber delay line before it is transduced into an electrical signal by 
    a photodetector; the electrical signal is filtered before driving the
    EOAM's electrical input, closing the feedback loop.

    The system is approximately linear when the EOAM is biased mid-fringe and the photodetector is 
    operated in its linear regime. In this case, the small-signal response of the open-loop system 
    is described by a frequency response $H_\t{OL}[\Omega]$ that relates the voltage at the EOAM's electrical input to the voltage at the photodetector's output. The
    system will oscillate around the frequency $\omega_{0}$ where the loop transmission is unity and the round-trip
    phase is an integer multiple of $2\pi$, i.e. $|H_{\text{OL}}[\omega_{0}]| = 1$ and
    $\arg H_\t{OL}[\omega_{0}] = 2\pi n$, for some integer $n$.

    Oscillation is initiated by positive feedback of any noise in the loop, resulting in a 
    self-sustained voltage oscillating at $\omega_{0}$ across the EOAM, which appears as modulated
    sidebands on the optical carrier.     
    The amplitude of the voltage and of the sidebands is determined by a balance between the 
    loop gain and the dominant nonlinearity in the system. 
    The same nonlinearity will reduce the small-signal open-loop gain at the oscillation frequency to unity
    such that $H_{\text{OL}}[\Omega] \rightarrow H_{\text{OL}}[\Omega]/|H_{\text{OL}}[\omega_{0}]|$.

    Gain saturation arises in the OEO primarily due to nonlinearities in the EOAM and the photodetector.
    In the EOAM, gain saturation arises from its sinusoidal response to RF voltages, which increases the amplitude of the transmitted optical sidebands at the expense of the optical carrier
    (see Supplementary Information). 
    The large-signal response of the photodetector's transimpedance amplifier can give rise 
    to both gain compression (from clipping) and phase distortion (from finite slew rate). 
    The gain and phase response of both the EOAM and the transimedance amplifier to an increasing 
    RF drive at the EOAM directly determine the steady-state behavior of the OEO's output power 
    and oscillation frequency.

    To quantify our OEO's steady-state gain saturation, we measure the its open-loop response as a function of RF drive power to the EOAM (as detailed in the Supplementary Information).
    \Cref{fig:classical_characterization}(a) shows this gain as a function of RF power, 
    and predicts that the OEO's gain will compress to unity with about 5 dBm of RF output power 
    at the EOAM. 
    Comparison with a theoretical model of the EOAM's nonlinear response (see Supplementary Information), 
    shows that the observed gain saturation of the loop is incompatible with the EOAM serving as 
    the sole saturating element. This suggests that the photodetector may be appreciably saturating as well. Lending further credence to this hypothesis, phase distortion also couples the RF output power and open-loop phase shift, shown in the measurements in the lower panel 
    of \cref{fig:classical_characterization}(a). Operating the OEO with an RF output power between 5 and 10 dBm minimizes this effect, which would otherwise result in an undesirable RF output 
    power-dependent frequency drift.

    Once nonlinearities have saturated the OEO's mean circulating and out-coupled fields, its
    closed loop response is given by
    \begin{equation}
        \label{eq:clTfDef}H_{\text{CL}}[\Omega] = \left(1 - \frac{H_{\text{OL}}[\Omega]}{|H_{\text{OL}}[\omega_{0}]|}\right
        )^{-1},
    \end{equation}
    which is shown in \cref{fig:classical_characterization}(b). 
    This measurement predicts that the OEO will oscillate at $9.52 \pm 0.01$ MHz.
    \Cref{fig:classical_characterization}(c) shows the measured spectrum of the OEO's RF output,
    which peaks at $\omega_0 \approx 2\pi \cdot 9.51$~MHz, consistent with the frequency predicted 
    by the closed loop response. 
    The power in this carrier, estimated by integrating the voltage spectrum in a 200 kHz band around it,
    is 6.6 dBm, consistent with the measured gain saturation. 

    \emph{Phase noise of the OEO output. } The fluctuations in the phase of the oscillating 
    RF carrier at $\omega_0$ determine the OEO's phase stability.
    In a quantum-noise-limited OEO, these phase fluctuations arise from quantum fluctuations in the optical field incident on the photodetector, which are fed back through the OEO loop.
    In reality, optical noises other than from quantum fluctuations of the optical field, and 
    electronic noise from the photodetector and electronics chain, also contribute,
    so that the power spectral density of the phase of the RF output is 
    (see Supplementary Information for a derivation)
    \begin{equation}
        \label{eq:oeoNoiseModel}
        \begin{split}
            \bar S_{\varphi \varphi}^{\text{out}}[\Omega] 
            \approx& \frac{ \left| H_{\text{CL}}[\omega_{0}+
            \Omega] \right|^{2}}{v_{\text{out}}^{2}}\\&\times \left(\lambda^2 \hbar \omega_\ell P_{0}
            \bar{S}^{\text{opt}}_{qq}[\omega_{0}] + 2\bar{S}^{\text{elec}}_{VV}[\omega_{0}] \right).
        \end{split}
    \end{equation}
    Here $S_{VV}^\t{elec}$ is the sum of all electronic noises referred to the photodetector's output
    voltage in the absence of optical power, and $\bar{S}_{qq}^\t{opt}$ is the optical noise due to
    fluctuations in the amplitude quadrature $\hat{q}$ of the field incident on the photodetector
    ($\lambda$ is the power to voltage response of the photodetector and electronics chain, and $P_0$ is the incident
    optical power at optical carrier frequency $\omega_\ell$).
    When the latter is dominated by quantum fluctuations, it is given by
    \begin{equation}\label{eq:oeoOpticalQN}
        \begin{split}
            \bar{S}^{\text{opt,qn}}_\t{qq}[\omega_{0}] =&(1-\eta_{\text{sqz}}^{2})(J_{0}(\beta)^{2}+ J_{1}(\beta)^{2}) \\&+ 2\eta_{\text{sqz}}
            ^{2}\left( J_{0}(\beta)^{2}\bar S_{qq}[\omega_{0}] + J_{1}(\beta)^{2}\bar S_{qq}[2\omega_{0}] \right).
        \end{split}
    \end{equation}
    Here we allow for the possibility of modifying the optical quantum fluctuations by injection of
    squeezed light before the photodetector, quantified by the transmission $\eta_\t{sqz}^2$ 
    of that path. The sinusoidal response of the EOAM to its RF drive results in the Bessel functions
    $J_i(\beta)$ with $\beta$ the modulation depth.

    In order for an OEO's phase noise to be well modeled by \cref{eq:oeoNoiseModel}, 
    all other noise contributions must lie well below those from quantum and electronics noise. 
    Our OEO relies on a combination of active stabilization and passive isolation to suppress other potential noise sources.
    For instance, we use a solid-state laser with free-running intensity noise at the quantum noise level at 9.5 MHz offset from the carrier. 
    Additionally, we actively stabilize the laser's long-term intensity drift with a low-bandwidth feedback loop to an optical attenuator.
    This prevents the OEO's mean frequency from drifting due to gain-phase coupling effects, discussed above.
    To limit the effect of fiber length fluctuations, we use a relatively short fiber delay line.
    Finally, using a relatively low-Q RF filter limits frequency pulling effects.
    Together, these design choices limit the influence of noises other than those from quantum and electrical fluctuations on the OEO's phase stability, such that it is modeled by \cref{eq:oeoNoiseModel}.

\begin{figure}
        \centering
        \includegraphics[width=1.0\columnwidth]{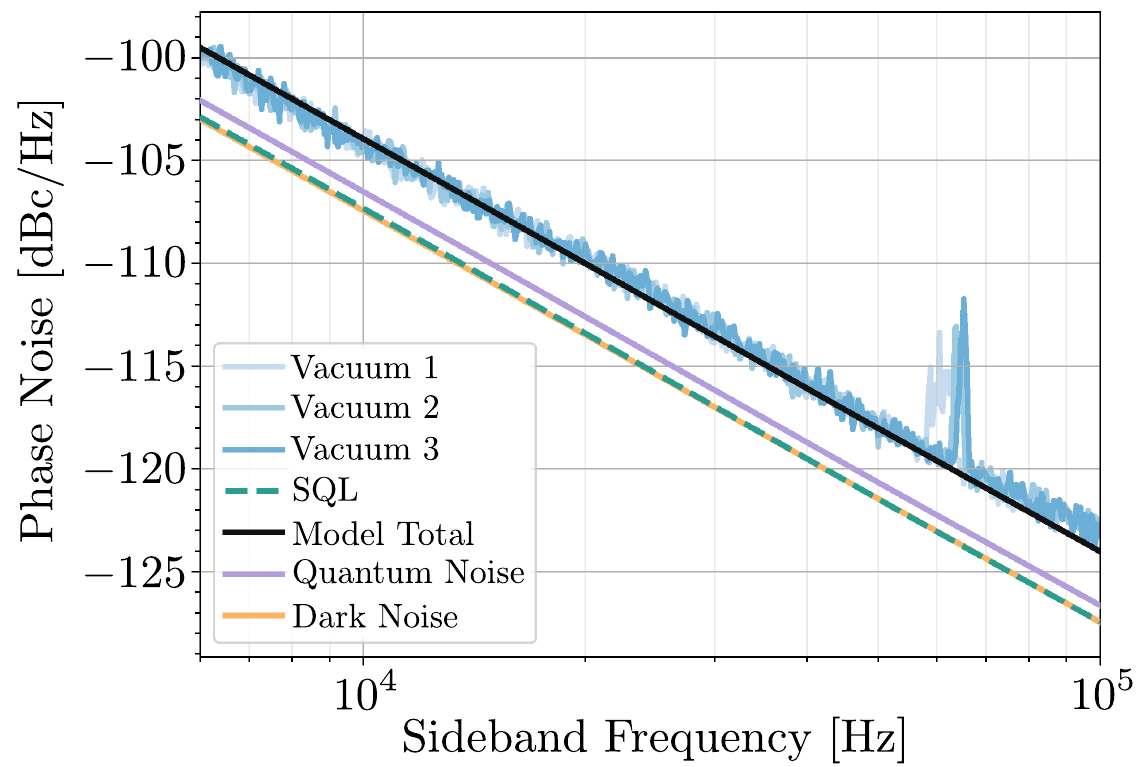}
        \caption{Opto-electronic oscillator phase noise. Comparison of our OEO's phase noise to the noise model in \cref{eq:oeoNoiseModel}.
        Vacuum 1,2, and 3 are taken sequentially with vacuum states on the squeezer port to verify that the oscillator's phase noise is stable over the duration of the measurements.}
        \label{fig:oeoNoiseExpVsModel}
    \end{figure}

    The key implication of \cref{eq:oeoNoiseModel} is 
    that a quantum-noise-limited OEO is realizable if the in-loop photodetector is 
    sufficiently low-noise that electronic noises become subdominant to optical quantum fluctuations 
    around the RF carrier frequency $\omega_0$.
    \Cref{fig:oeoNoiseExpVsModel} shows the OEO's measured phase noise compared with the model in
    \cref{eq:oeoNoiseModel} with no squeezed light injection. 
    The model is determined from independent measurements of the response of the feedback loop 
    elements as well as the optical power ($P_0 \approx 64\, \mu\text{W}$) and has no free parameters.
    Quantum noise dominates over electronic noise by about $0.8\, \t{dB}$ 
    across the measured frequencies, and the measurement agrees with the model. 
    In this sense, 
    our OEO is quantum-noise-limited.

    \begin{figure*}[!t]
        \centering
        \includegraphics[width=1.0\textwidth]{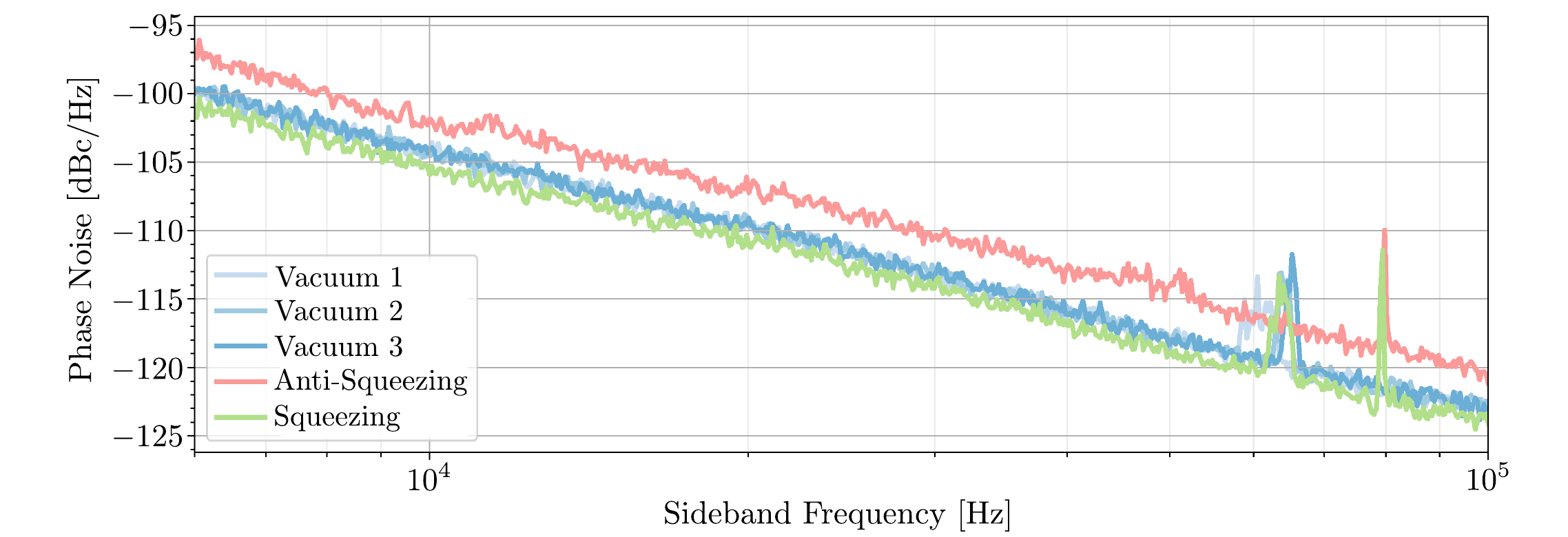}
        \caption{Quantum-enhanced opto-electronic oscillator phase noise. By injecting squeezed light, we suppress the
        OEO's phase noise spectrum by 1.0 dB. By injecting anti-squeezed light, we can increase its phase noise
        spectrum by 2.8 dB. The trace ``Vacuum 1'' was taken before measuring squeezing, ``Vacuum 2'' was
        taken after measuring squeezing and before measuring anti-squeezing, and ``Vacuum 3'' was taken after measuring anti-squeezing.
        These vacuum traces are the same as those in \cref{fig:oeoNoiseExpVsModel}.
        This ensures that the oscillator's phase noise did not drift over the duration of the measurement and that any
        phase noise suppression or amplification is due to squeezed light injection. }
        \label{fig:oeoQuantumEnhancement}
    \end{figure*}

    The degree to which a feedback oscillator's phase stability is limited by quantum noise 
    is quantified by its proximity to the standard quantum limit (SQL).
    For an OEO, we define the SQL to be the ideal phase stability achievable in the absence of any quantum enhancements, i.e. (see also Ref. \cite{YaoMal96a})
    \begin{equation}
        \label{eq:oeoSql}\bar{S}_{\varphi \varphi}^{\text{SQL}}[\Omega] = \frac{\hbar \omega_{\ell}}{P_{0}}|H_{\text{CL}}
        [\omega_{0}+ \Omega]|^{2}\left(\frac{J_{0}(\beta)^{2}+ J_{1}(\beta)^{2}}{J_{1}(2\beta)^{2}}\right).
    \end{equation}
    Note that the SQL so-defined is achievable only when the photodetector's quantum efficiency 
    is unity (see Supplementary Information), and can therefore be distinct from the 
    quantum noise contribution. Indeed in \cref{fig:oeoNoiseExpVsModel}, the quantum noise lies above the SQL due to our photodetector's quantum efficiency of 83\%.

    We see from \cref{fig:oeoNoiseExpVsModel} that our OEO lies within 3.5 dB of the SQL.
    This close proximity to the SQL opens the possibility of
    engineering the OEO's quantum noise and the demonstration of quantum enhancement in its phase
    stability for the first time.

    \emph{Quantum-enhanced phase stability. }
    We inject squeezed light from
    a degenerate optical parametric amplifier (OPA) into the OEO, 
    such that the optical field measured by 
    the OEO's photodetector is a displaced squeezed state rather than a displaced vacuum state.
    The squeezed state produced by the OPA has quantum fluctuations suppressed below the vacuum noise level along one quadrature
    axis, and fluctuations increased above this level along the opposite quadrature axis. The degree of quantum noise increase
    or suppression in the OEO depends on the relative orientation between the squeezed field and the laser beam's amplitude quadrature (measured by the photodetector). 
    We deterministically set this orientation 
    using an auxiliary coherent control field \cite{Vahlbruch06,Ganapathy23} with feedback to the phase of the OEO's LO field and an error signal derived from an out-of-loop photodetector as depicted in \cref{fig:simple_schematic} (see Supplementary Information for further details). 

    \Cref{fig:oeoQuantumEnhancement} shows the OEO's phase noise spectrum with squeezed light injection,
    compared against the case with no squeezed light injection.
    When the squeezing is oriented to suppress the OEO's phase noise, we observe a broadband reduction
    in phase noise by about 1.0 dB between 6 kHz and 100 kHz offset from the 9.51 MHz carrier.
    To our knowledge, this is the first experimental demonstration of a feedback
    oscillator with phase stability enhanced through quantum engineering.

    As an independent verification that the observed phase-noise suppression is due to squeezed light,
    we tune the relative orientation of the squeezed field to instead amplify the OEO's
    phase noise by 2.8 dB due to anti-squeezing.
    This also allows an independent characterization of the losses along the squeezed light path
    as well as the level of squeezing at the photodetector. 
    We find that the losses, including imperfect quantum efficiency, are roughly 45\%, and that the generated squeezing level is roughly 5 dB at 9.5 MHz. The losses arise
    from a combination of imperfect coupling from the free-space OPA to fiber, optical losses in the fiber components, and an
    imperfect photodetector quantum efficiency of $0.83$. 
    Reducing these losses and increasing the dark-noise clearance would
    enable increased quantum enhancement.

    In addition to offering the potential to improve an OEO's phase stability though squeezed light injection, 
    coupling a squeezed light source to an OEO also introduces new potential noise couplings that can reduce the OEO's phase stability if not mitigated.
    In particular, we found it essential to use a pair of Faraday isolators to  achieve at least $60$ dB of optical isolation between our OPA and the OEO. We believe this high level of isolation serves to suppress backscattered light and a parasitic etalon formed between the OEO and the linear OPA's input-output coupling mirror, and that future realizations of quantum-enhanced feedback oscillators will need to use similar isolation strategies to ensure that their quantum sources do not destabilize the oscillator due to such parasitic effects.

    \textit{Conclusion.} We have constructed an opto-electronic oscillator with quantum-enhanced phase stability.

    Realizing an oscillator with quantum-enhanced phase stability imposes two fundamental
    requirements. 
    First, the oscillator's phase fluctuations must be dominated by quantum noise, rather than 
    classical noises or environmental drifts; prior experiments with feedback oscillators have demonstrated 
    this \cite{Sieg68,Fritschel1989,Benmessai2008,Breeze18}. 
    The second requirement is that the quantum noise must be engineered to suppress phase 
    fluctuations.
    To our knowledge, no previous experiment has achieved quantum-enhanced phase stability.

    The phase stability of a feedback oscillator in the absence of quantum enhancement is 
    limited by an SQL \cite{Loughlin23}.
    In the case of lasers and masers, this SQL is the famous Schawlow-Townes limit.
    Quantum engineering techniques have been proposed to evade the SQL \cite{Loughlin23,Loughlin26}.
    The current experiment demonstrates the principle of such quantum enhancement; with further 
    improvements in reduction of losses and dark-noise, we expect that an OEO can be made to 
    operate at and beyond the SQL. Similar techniques may then be applied to other feedback
    oscillators such as lasers and masers, enabling a new class of coherent sources 
    with quantum-enhanced phase stability.\\

    \emph{Acknowledgments. }We gratefully acknowledge the support of the 
    National Science Foundation (NSF) 
    through the LIGO operations cooperative agreement PHY-18671764464 and NSF award 2308969.
    This work was made possible through the support of the Enrico Fermi Fellowships led by the Center for Spacetime and the Quantum, 
    and supported by Grant ID \#63132 from the John Templeton Foundation. 
    D.M.D is supported by the NSF Graduate Research Fellowship Program (award 2141064).
    V.S. is supported in part by an NSF CAREER award (PHY–2441238), 
    Department of Energy QuantISED 2.0 grant (DE-SC0026164),  
    and by the Gordon and Betty Moore Foundation (grant GBMF13780).
    We thank the MIT Radio Society for lending us the Agilent PXA signal analyzer used in this work.

    % ---------- Bibliography ----------
    \bibliographystyle{apsrev4-2}
    \bibliography{main}

    \appendix
    \onecolumngrid

    \section{Amplitude and phase modulation quadratures}

    The magnitude of a quantized electromagnetic field with angular frequency $\omega_{\ell}$, normalized to its mode profile
    with cross-sectional area $\mathcal{A}$ is given by \cite{Blow90,Danilishin2012}
    \begin{equation}
        \hat{E}(t) = E_{0}(\omega_{\ell}) \left[ \hat{q}(t) \cos(\omega_{\ell}t) + \hat{p}(t) \sin(\omega_{\ell}t) \right
        ],
    \end{equation}
    with the normalization constant $E_{0}(\omega_{\ell}) := \sqrt{4\pi\hbar \omega_{\ell}/(\mathcal{A}c)}$. The
    operators $\hat{q}$ and $\hat{p}$ obey the time-domain commutation relation
    $[\hat{q}(t),\hat{p}^{\dagger}(t^{\prime})] = i\delta(t-t^{\prime})$. These are the quantized optical amplitude and phase
    modulation quadratures with respect to a laser beam at frequency $\omega_{\ell}$.

    If the field has a mean value $\langle \hat{q}\rangle = \bar q$ such that $\delta\hat{q}$ and $\delta\hat{p}$ represent
    fluctuating quantities with mean values of zero, then the electromagnetic field is
    \begin{equation}
        \hat{E}(t) = E_{0}(\omega_{\ell}) \left[\bar q + \delta\hat{q}(t) \right] \cos\left[ \omega_{\ell}t - \hat{p}(t) /
        \bar q \right].
    \end{equation}
    From this equation, we see that $\delta\hat{q}(t)$ and $\delta\hat{p}(t)/\bar q$ represent intensity and phase fluctuations
    of the electromagnetic field.

    \section{Fourier transform conventions}

    We define the Fourier transform of a Heisenberg frame operator $\hat X(t)$ by
    \begin{equation}
        \hat{X}[\Omega] := \int_{-\infty}^{\infty}dt \, e^{i\Omega t}\hat{X}(t),
    \end{equation}
    which has the inverse
    \begin{equation}
        \hat{X}(t) = \int_{-\infty}^{\infty}\frac{d\Omega}{2\pi}\, e^{-i\Omega t}\hat{X}[\Omega].
    \end{equation}

    We call an operator $\hat X$ weak-stationary if it satisfies $\langle \hat X(t) \hat X(t^{\prime}) \rangle = \langle \hat
    X(t-t^{\prime}) \hat X(0) \rangle$. We define the symmetrized auto-correlation of a weak-stationary operator by
    \cite{Clerk10}
    \begin{equation}
        \bar S_{XX}(t) = \frac{1}{2}\langle \{ \hat X^{\dagger}(t), \hat X(0) \} \rangle,
    \end{equation}
    and we define the symmetrized power spectral density of an operator to be the Fourier transform of its symmetrized
    autocorrelation function:
    \begin{equation}
        \bar S_{XX}[\Omega] = \int_{-\infty}^{\infty}\frac{1}{2}\langle \{ \hat X[\Omega], \hat X[\Omega^{\prime}]^{\dagger}
        \} \rangle \frac{d\Omega^{\prime}}{2\pi}.
    \end{equation}

    \section{Up- and down-converted noise}
    \label{sec:up_and_down_converted_noise}

    The OEO's photodetector is a nonlinear square-law detector that allows noise to be shifted between different frequency
    bands. In this section, we discuss how to analyze such up- and down-converted noise with respect to a carrier at
    frequency $\omega_{0}$.

    Consider a system with a coherent tone at RF frequency $\omega_{0}$ with mean amplitude $a_{0}$, amplitude fluctuations
    $\delta a(t)$ and phase fluctuations $\delta \phi(t)$:
    \begin{equation}
        x(t) = (a_{0}+ \delta a)\cos(\omega_{0}t) - a_{0}\delta \phi (t) \sin(\omega_{0}t).
    \end{equation}
    Let the signal $y(t)$ be $x(t) + \delta n(t)$, where $\delta n (t)$ represents additional noise. To see how this
    additional noise contributes to the new signal's amplitude and phase noise, we define the quadrature envelopes of $\delta
    n(t)$ about the carrier frequency $\omega_{0}$ by
    \begin{equation}
        \begin{split}
            \delta A_{\text{n}}(t)&= \text{LPF}[ 2 \delta n(t) \cos (\omega_{0}t) ] \\ \delta \Phi_{\text{n}}(t)&= \text{LPF}
            [ -2 \delta n(t) \sin (\omega_{0}t) ].
        \end{split}
    \end{equation}
    Here $\text{LPF}[\cdot ]$ represents the time-domain response of a low-pass filter with a passband width of less than
    $\omega_{0}$, but which is otherwise left arbitrary for now.

    In terms of these quadratures, $\delta n(t)$ is
    \begin{equation}
        \delta n(t) = \delta A_{\text{n}}(t) \cos(\omega_{0}t) - \delta \Phi_{\text{n}}(t) \sin(\omega_{0}t),
    \end{equation}
    and we can now write $y(t)$ as
    \begin{equation}
        \begin{split}
            y(t) =&[a_{0}+ \delta a(t) + \delta A_{\text{n}}(t)]\cos(\omega_{0}t) - [a_{0}\delta \phi (t) + \delta \Phi
            _{\text{n}}(t)] \sin(\omega_{0}t).
        \end{split}
    \end{equation}
    That is, $\delta A_{\text{n}}$ and $\delta \Phi_{\text{n}}$ are additive quadrature envelopes with the same units as
    $x(t)$. The dimensionless phase fluctuation contributed by $\delta n(t)$ is $\delta \varphi_{\text{n}}(t) = \delta \Phi
    _{\text{n}}(t)/a_{0}$. In general, for a signal with mean carrier amplitude $a_{i}$, we will use the notation that
    $\delta \Phi_{i}$ denotes the signal's phase-quadrature fluctuations, with the same units as $a_{i}$, and
    $\delta\varphi_{i}$ denotes the signal's phase fluctuations, with units of angular frequency.

    We will find the following relation helpful below:
    \begin{equation}
        \begin{split}
            \int_{-\infty}^{\infty}dt e^{i \Omega t}\text{LPF}[\sin(\omega_{0}(t-\tau)) \delta x(t-\tau^{\prime})] = 
            \frac{-i e^{i\Omega \tau^\prime}H_{\text{lpf}}[\Omega]}{2}\Big( e^{i\omega_0 (\tau^\prime - \tau)}\delta
            x[\Omega + \omega_{0}] - e^{-i\omega_0 (\tau^\prime - \tau)}\delta x[\Omega
            - \omega_{0}] \Big).
        \end{split}
    \end{equation}
    Here, $H_{\text{lpf}}[\Omega]$ is the transfer function of the arbitrary low pass filter.

    \section{Quantum optical model of an opto-electronic oscillator}

    In this section, we derive a noise model for our opto-electronic oscillator in a manner that makes it clear how quantum
    fluctuations emerge in the device and how they may be suppressed through squeezed light injection. We begin by
    modelling the coupled equations of motion that describe the OEO's time-domain response.

    We can model the OEO's EOAM as a variable beam splitter with amplitude transmissivity
    $\cos(\pi v_{\text{mod}}(t)/(2 v_{\pi}) +\phi_{\text{bias}})$. We let $\hat{x}\in \{ \hat q, \hat p \}$ denote an arbitrary
    quadrature of the optical field, which is convenient for analyzing optical elements that do not distinguish between
    the amplitude and phase quadratures. The quadratures of the field transmitted through the EOAM,
    $\hat x_{\text{eoam}}(t)$, are related to those into the EOAM, $\hat x_{\text{in}}(t)$, and those of a vacuum mode $\hat
    x_{\text{eoam}}^{\text{vac}}(t)$ by
    \begin{equation}
        \begin{split}
            \hat x_{\text{eoam}}(t) =&\cos \left(\pi\frac{v_{\text{mod}}(t)}{2v_{\pi}}+\phi_{\text{bias}}\right) \hat x_{\text{in}}
            (t) + \sin \left(\pi\frac{v_{\text{mod}}(t)}{2v_{\pi}}+\phi_{\text{bias}}\right) \hat x_{\text{eoam}}^{\text{vac}}
            (t).
        \end{split}
    \end{equation}
    The quadratures after the fiber delay line, $\hat x_{\text{fiber}}$ are delayed by time $\tau$ with respect to those
    at its input such that
    \begin{equation}
        \hat x_{\text{fiber}}(t) = \hat x_{\text{eoam}}(t-\tau).
    \end{equation}
    These fields also experience imperfect optical power transmission $0 < \eta_{\ell}^{2}< 1$ through the fiber
    components, as well as an effective loss due to imperfect quantum efficiency in the photodetector. We model these
    effects by a passive beam splitter between the optical fiber and the photodetector, such that the photodetector measures
    the field given by
    \begin{equation}
        \hat x_{\text{pd}}(t) = \eta \, \hat x_{\text{fiber}}(t) + \sqrt{1-\eta^{2}}\, \hat x_{\text{loss}}^{\text{vac}}(
        t),
    \end{equation}
    with unit quantum efficiency. Here, $\eta = \eta_{\text{qe}}\eta_{\ell}$ is the effective amplitude transmissivity resulting
    from optical losses and the detector's quantum efficiency.
    $\hat x_{\text{loss}}^{\text{vac}}(t)$ represents vacuum noise in-coupled from optical losses and imperfect quantum
    efficiency.

    In terms of the input field quadratures, the field measured by the photodetector is
    \begin{equation}
        \label{eq:xinToxpd}
        \begin{split}
            \hat x_{\text{pd}}(t) =&\eta \cos \left(\pi\frac{v_{\text{mod}}(t-\tau)}{2 v_{\pi}}+\phi_{\text{bias}}\right)
            \hat x_{\text{in}}(t-\tau) + \eta\sin \left(\pi\frac{v_{\text{mod}}(t-\tau)}{2 v_{\pi}}+\phi_{\text{bias}}
            \right) \hat x_{\text{eoam}}^{\text{vac}}(t-\tau) + \sqrt{1-\eta^{2}}\, \hat x_{\text{loss}}^{\text{vac}}(
            t).
        \end{split}
    \end{equation}

    We denote the mean value of an operator by $\bar{X}:= \langle \hat{X}\rangle$, where the average is taken with respect
    to the quantum state, $\hat \rho$, and the remaining fluctuations are denoted by $\delta\hat x := \hat X - \bar X$ such
    that $\hat X = \bar X + \delta \hat X$. For amplitude and phase quadratures $\hat q$ and $\hat p$, we have $\hat q = \bar
    q + \delta \hat q$ and $\hat p = \delta \hat p$. The photodetector's photocurrent is then
    \begin{equation}
        \label{eq:photocurrent}
        \begin{split}
            i_{\text{pd}}(t) =&\, \frac{q_{e}}{2}\left( \hat q_{\text{pd}}(t) - i \hat p_{\text{pd}}(t) \right)\left( \hat
            q_{\text{pd}}(t) + i \hat p_{\text{pd}}(t) \right) \\ =&\, q_{e}\left( \bar q_{\text{pd}}(t)^{2}/2 + \bar q_{\text{pd}}
            (t) \delta \hat q_{\text{pd}}(t) + \mathcal{O}(\delta \hat x_{\text{pd}}^{2})\right),
        \end{split}
    \end{equation}
    where $q_{e}$ is the charge of an electron. We can also relate the mean photocurrent to the power incident on the photodetector
    via $\bar q_{\text{pd}}(t)^{2}= 2 P_{\text{pd}}(t) / (\hbar \omega_{\ell})$ \cite{sudhir2017}. Defining the
    photodetector's responsivity, $R := q_{e}/(\hbar \omega_{\ell})$, we also have
    \begin{equation}
        \label{eq:photocurrentFromPower}\bar i_{\text{pd}}(t) = R P_{\text{pd}}(t),
    \end{equation}
    which will be useful in the mean-field analysis below.

    We assume the photodetector has a broadband transimpedance gain $g$ (in practice that the photodetector's bandwidth is
    much larger than the oscillation frequency), in which case it produces a voltage
    \begin{equation}
        \label{eq:transimpedance}v_{\text{pd}}(t) = g \, i_{\text{pd}}(t).
    \end{equation}
    This in turn determines the modulator's voltage, $v_{\text{mod}}(t)$ via
    \begin{equation}
        \label{eq:filter}v_{\text{mod}}(t) = \int_{-\infty}^{\infty}H_{\text{elec}}(t-T)v_{\text{pd}}(T) d T,
    \end{equation}
    where $H_{\text{elec}}(t)$ is the time domain impulse response of the electronics chain containing the bandpass
    filter and RF splitter.

    \subsection{Mean-field response}

    To understand the oscillator's phase and frequency fluctuations, we separate its dynamics into a nonlinear mean-field
    response and a linearized response about this mean field. Neglecting fluctuation terms, we have (from \cref{eq:xinToxpd,eq:photocurrent,eq:photocurrentFromPower,eq:transimpedance,eq:filter})
    \begin{equation}
        \label{eq:oscCondition}
        \begin{split}
            \bar v_{\text{mod}}(t) =&g \eta^{2}R P_{\text{in}}\int_{-\infty}^{\infty}H_{\text{elec}}(t-T) \cos^{2}
            \left(\pi\frac{\bar v_{\text{mod}}(T-\tau)}{2 v_{\pi}}+\phi_{\text{bias}}\right) dT.
        \end{split}
    \end{equation}
    Since this system is an oscillator, we expect the mean-field to be a sinusoid at some frequency $\omega_{0}$. In
    light of this, we take ansatz $\bar v_{\text{mod}}(t) = v_{0}\cos(\omega_{0}t )$. We also assume that the filter
    rejects oscillations at frequencies $n \omega_{0}$ for integer $n \neq 1$ and that the modulator is biased about mid-fringe
    such that $\phi_{\text{bias}}= \pi/4$. We then have
    \begin{equation}
        \begin{split}
            \bar v_{\text{mod}}(t) =&\frac{g \eta^{2}R P_{\text{in}}}{2}\int_{-\infty}^{\infty}H_{\text{elec}}(t-T) 
            \left[ 1 - \sin\left(\pi\frac{\bar v_{\text{mod}}(T-\tau)}{v_{\pi}}\right) \right] dT.
        \end{split}
    \end{equation}
    Defining $\beta := \pi v_{0}/v_{\pi}$ and $\Theta(T) := \omega_{0}(T-\tau)$, we expand
    \begin{equation}
        \sin(\beta \cos \Theta) = 2 \sum_{m=0}^{\infty}(-1)^{m}J_{2m+1}(\beta) \cos[(2m+1)\Theta].
    \end{equation}
    Neglecting the DC response and harmonics at frequencies $n \omega_{0}$ with $n \neq 1$, we find the steady-state oscillation
    condition
    \begin{equation}
        \label{eq:expOscCond}
        \begin{split}
            v_{0}\cos(\omega_{0}t ) = g \eta^{2}R P_{\text{in}}J_{1}\left( \frac{\pi v_{0}}{v_{\pi}}\right) |H_{\text{elec}}
            [\omega_{0}]| \cos(\omega_{0}t -\omega_{0}\tau + \phi_{\text{elec}}+ \pi),
        \end{split}
    \end{equation}
    where $J_{n}$ is the $n^{\text{th}}$ Bessel function of the first kind and $|H_{\text{elec}}[\omega_{0}]|$ and $\phi_{\text{elec}}
    := \text{arg}(H_{\text{elec}}[\omega_{0}])$ are the magnitude and phase of the filter response at frequency $\omega_{0}$.

    For small signals $\pi v_{0}/v_{\pi}\ll 1$ such that $J_{1}(\pi v_{0}/v_{\pi}) = \pi v_{0}/(2 v_{\pi}) + \mathcal{O}(
    (v_{0}/v_{\pi})^{3})$. In this case, we can read the OEO's unsaturated open-loop transfer function off from \cref{eq:expOscCond},
    and find it to be
    \begin{equation}
        \label{eq:olTfApp}
        \begin{split}
            H_{\text{OL}}[\Omega] = - \frac{\pi g H_{\text{elec}}[\Omega] \eta^{2}R P_{\text{in}}e^{i \Omega\tau}}{2 v_{\pi}}
        \end{split}
    \end{equation}

    From \cref{eq:expOscCond,eq:olTfApp} we see that oscillation occurs at frequency $\omega_{0}$ when $\arg(H_{\text{OL}}
    [\omega_{0}]) = 2\pi$ and $|H_{\text{OL}}[\omega_{0}]| \geq 1$. Generally, we will have $|H_{\text{OL}}[\omega_{0}]| >
    1$ initially, and saturation effects will reduce the magnitude of $H_{\text{OL}}$ until
    $|H_{\text{OL}}[\omega_{0}]| = 1$.

    If saturation occurs primarily due to the modulator's nonlinearity, we can model the OEO's saturating behavior analytically.
    For oscillation amplitudes satisfying $\pi v_{0}/v_{\pi}\ll 1$, we solve \cref{eq:expOscCond} analytically using $J_{1}
    (\beta) = \beta/2 - \beta^{3}/16 + \mathcal{O}(\beta^{5})$, from which we find that $v_{0}$ is given by
    \begin{equation}
        \label{eq:oscAmpSol}v_{0}=
        \begin{cases}
            0                                                                 & \text{ for }0 \leq |H_{\text{OL}}[\omega_{0}]| < 1 \\
            \frac{2 v_\pi}{\pi}\sqrt{2(1- \frac{1}{|H_\text{OL}[\omega_0]|})} & \text{ for }|H_{\text{OL}}[\omega_{0}]| \geq 1
        \end{cases}
    \end{equation}
    More generally, we can solve the transcendental equation $v_{0}= g \eta^{2}R P_{\text{in}}J_{1}\left( \frac{\pi v_{0}}{v_{\pi}}
    \right) |H_{\text{elec}}[\omega_{0}]|$ numerically as long as the saturation is due to the modulator. As discussed
    in the main, in our OEO both the photodetector and modulator saturate, so the saturation dynamics are not captured
    by \cref{eq:oscAmpSol}.

    \subsection{Fluctuations around the mean field}

    We now analyze the system's fluctuations about its mean field response. We consider two sources of these
    fluctuations, quantum noise in the form of quadrature fluctuations $\delta \hat x$, and electronics dark noise
    $\delta v_{\text{d}}$, primarily due to the photodetector. These noise sources lead to fluctuations in the phase of
    the oscillator's output voltage $v_{\text{out}}(t) = C v_{\text{mod}}(t)$, where $C$ is a proportionality constant between
    the oscillator's voltage across the modulator and out-coupled voltage. Physically, this constant is determined by the
    RF splitter used to out-couple an electronic signal from the oscillator. In the OEO discussed in this manuscript, we have
    $C = 1$.

    From \cref{eq:xinToxpd}, and using $\phi_{\text{bias}}= \pi/4$, we can derive an approximate form of the amplitude quadrature
    seen by the photodetector. We write the amplitude and phase fluctuations of $v_{\text{mod}}$ as
    $\delta A_{\text{mod}}$ and $\delta \varphi_{\text{mod}}$ such that
    \begin{equation}
        v_{\text{mod}}(t) = (v_{0}+ \delta A_{\text{mod}}(t)) \cos(\omega_{0}t + \delta \varphi_{\text{mod}}(t)) \approx v
        _{0}\cos(\omega_{0}t ) + \delta A_{\text{mod}}(t) \cos(\omega_{0}t ) - v_{0}\delta \varphi_{\text{mod}}(t) \sin (
        \omega_{0}t ).
    \end{equation}
    Defining $J_{n}:= J_{n}(\pi v_{0}/(2 v_{\pi}))$, and retaining only terms linear in $\delta A_{\text{mod}}$,
    $\delta \varphi_{\text{mod}}$, and the quadrature fluctuations together with harmonics at frequencies $0$ and $\omega
    _{0}$, we then have
    \begin{equation}
        \begin{split}
            \hat q_{\text{pd}}(t) \approx&\frac{\eta}{\sqrt{2}}\bigg[ J_{0}- 2 J_{1}\cos(\omega_{0}(t - \tau)) \bigg] \bar
            q_{\text{in}}\\&+ \frac{\eta}{\sqrt{2}}\bigg[ - \frac{\pi \delta A_{\text{mod}}(t-\tau)}{2 v_{\pi}}J_{1}- \frac{\pi
            \delta A_{\text{mod}}(t-\tau)}{2 v_{\pi}}J_{0}\cos(\omega_{0}(t - \tau)) + \frac{\pi v_{0}}{2 v_{\pi}}J_{0}\delta
            \varphi_{\text{mod}}(t - \tau) \sin(\omega_{0}(t - \tau)) \bigg] \bar q_{\text{in}}\\&+ \frac{\eta}{\sqrt{2}}
            \bigg[ J_{0}- 2 J_{1}\cos(\omega_{0}(t - \tau)) \bigg] \delta \hat q_{\text{in}}(t-\tau) \\&+ \frac{\eta}{\sqrt{2}}
            \bigg[ J_{0}+ 2 J_{1}\cos(\omega_{0}(t - \tau)) \bigg] \delta \hat q_{\text{eoam}}^{\text{vac}}(t-\tau) \\&+ \sqrt{1-\eta^{2}}
            \, \delta \hat q_{\text{loss}}^{\text{vac}}(t),
        \end{split}
    \end{equation}
    where we have neglected terms that are second-order in the fluctuations and Bessel functions of higher order than
    $J_{1}$.

    The photodetector voltage is proportional to the square of this quantity. It has a mean value given by
    \begin{equation}
        \bar v_{\text{pd}}(t) = \frac{g q_{e}\eta^{2}\bar q_{\text{in}}^{2}}{4}\bigg[ J_{0}- 2 J_{1}\cos(\omega_{0}( t - \tau
        )) \bigg]^{2},
    \end{equation}
    with a carrier-frequency component $- g q_{e}\eta^{2}\bar q_{\text{in}}^{2}J_{0}J_{1}\cos(\omega_{0}(t-\tau))$, and
    fluctuations given by
    \begin{equation}
        \begin{split}
            \delta v_{\text{pd}}(t) =&- \frac{\pi g q_{e}\eta^{2}\bar q_{\text{in}}^{2}}{4 v_{\pi}}\bigg[ (J_{0}^{2}- 2 J
            _{1}^{2}) \cos(\omega_{0}(t - \tau)) - J_{0}J_{1}\cos(2 \omega_{0}(t - \tau)) \bigg] \delta A_{\text{mod}}(t-
            \tau) \\&+ \frac{\pi v_{0}g q_{e}\eta^{2}\bar q_{\text{in}}^{2}}{4 v_{\pi}}\bigg[ J_{0}^{2}\sin(\omega_{0}(t -
            \tau)) - J_{0}J_{1}\sin(2 \omega_{0}(t - \tau)) \bigg]\, \delta \varphi_{\text{mod}}(t - \tau) \\&+ \frac{g
            q_{e}\eta^{2}\bar q_{\text{in}}}{2}\bigg[ J_{0}^{2}+ 2 J_{1}^{2}- 4 J_{0}J_{1}\cos(\omega_{0}(t - \tau)) + 2 J
            _{1}^{2}\cos(2\omega_{0}(t - \tau)) \bigg] \delta \hat q_{\text{in}}(t-\tau) \\&+ \frac{g q_{e}\eta^{2}\bar
            q_{\text{in}}}{2}\bigg[ J_{0}^{2}- 2 J_{1}^{2}- 2 J_{1}^{2}\cos(2\omega_{0}(t - \tau) ) \bigg] \delta \hat q_{\text{eoam}}
            ^{\text{vac}}(t-\tau) \\&+ \eta g q_{e}\bar q_{\text{in}}\sqrt{\frac{1-\eta^{2}}{2 }}\bigg[ J_{0}- 2 J_{1}\cos
            (\omega_{0}(t - \tau)) \bigg] \, \delta \hat q_{\text{loss}}^{\text{vac}}(t) \\&+ \delta v_{\text{dark}}(t).
        \end{split}
    \end{equation}
    Here, $\delta v_{\text{dark}}(t)$ is the photodetector's dark noise voltage. This is the only term present in the
    absence of an optical signal, when $\bar q_{\text{in}}= 0$. In our OEO, this is the dominant noise source besides quantum
    noise.
    We observe that the terms $\cos(2 \omega_{0}(t - \tau)) \, \delta A_{\text{mod}}(t-\tau)$ and
    $\sin(2 \omega_{0}(t - \tau)) \, \delta \varphi_{\text{mod}}(t - \tau)$ are centered around $2 \omega_{0}$ and have
    noise bandwidths of less than $\omega_{0}$ since these fluctuations in the carrier field are suppressed by the RF
    filter centered at frequency $\omega_{0}$, which has a bandwidth of less than $\omega_{0}$. As a result, we neglect these
    modulation terms, while retaining the remaining $2 \omega_{0}$ contributions multiplying broadband quantum noise. We now
    have
    \begin{equation}
        \begin{split}
            \delta v_{\text{pd}}(t) =&- \frac{\pi g q_{e}\eta^{2}\bar q_{\text{in}}^{2}}{4 v_{\pi}}(J_{0}^{2}- 2 J_{1}^{2}
            ) \cos(\omega_{0}(t - \tau)) \, \delta A_{\text{mod}}(t-\tau) \\&+ \frac{\pi v_{0}g q_{e}\eta^{2}\bar q_{\text{in}}^{2}}{4
            v_{\pi}}J_{0}^{2}\sin(\omega_{0}(t - \tau)) \, \delta \varphi_{\text{mod}}(t - \tau) \\&+ \frac{g q_{e}\eta^{2}\bar
            q_{\text{in}}}{2}\bigg[ J_{0}^{2}+ 2 J_{1}^{2}- 4 J_{0}J_{1}\cos(\omega_{0}(t - \tau)) + 2 J_{1}^{2}\cos(2\omega
            _{0}(t - \tau)) \bigg] \delta \hat q_{\text{in}}(t-\tau) \\&+ \frac{g q_{e}\eta^{2}\bar q_{\text{in}}}{2}\bigg
            [ J_{0}^{2}- 2 J_{1}^{2}- 2 J_{1}^{2}\cos(2\omega_{0}(t - \tau) ) \bigg] \delta \hat q_{\text{eoam}}^{\text{vac}}
            (t-\tau) \\&+ \eta g q_{e}\bar q_{\text{in}}\sqrt{\frac{1-\eta^{2}}{2 }}\bigg[ J_{0}- 2 J_{1}\cos(\omega_{0}(
            t - \tau)) \bigg] \, \delta \hat q_{\text{loss}}^{\text{vac}}(t) \\&+ \delta v_{\text{dark}}(t).
        \end{split}
    \end{equation}
    We define the total quantum noise contribution to be
    \begin{equation}
        \label{eq:deltaQqnDef}
        \begin{split}
            \delta \hat{q}_{\text{qn}}(t) :=&\bigg[ J_{0}^{2}+ 2 J_{1}^{2}- 4 J_{0}J_{1}\cos(\omega_{0}(t - \tau)) + 2 J_{1}
            ^{2}\cos(2\omega_{0}(t - \tau)) \bigg] \delta \hat q_{\text{in}}(t-\tau) \\&+ \bigg[ J_{0}^{2}- 2 J_{1}^{2}- 2
            J_{1}^{2}\cos(2\omega_{0}(t - \tau)) \bigg] \delta \hat q_{\text{eoam}}^{\text{vac}}(t-\tau) \\&+ \sqrt{\frac{2(1-\eta^{2})}{
            \eta^{2}}}\bigg[ J_{0}- 2 J_{1}\cos(\omega_{0}(t - \tau)) \bigg] \, \delta \hat q_{\text{loss}}^{\text{vac}}(
            t).
        \end{split}
    \end{equation}
    As in the discussion of up- and down-converted noise above, we define
    modulation quadratures for up- and down-converted quantum noise using the same delayed carrier phase that appears in the
    feedback signal:
    \begin{equation}
        \label{eq:deltaPQqnDef}
        \begin{split}
            \delta \hat{A}_{\text{qn}}(t)&:= \text{LPF}\left[ 2 \, \delta \hat{q}_{\text{qn}}(t) \cos(\omega_{0}(t-\tau) )
            \right] \\ \delta \hat{\Phi}_{\text{qn}}(t)&:= \text{LPF}\left[ -2 \, \delta \hat{q}_{\text{qn}}(t) \sin(\omega
            _{0}(t-\tau)) \right],
        \end{split}
    \end{equation}
    Similarly, we define
    \begin{equation}
        \begin{split}
            \delta A_{\text{dark}}(t)&:= \text{LPF}\left[ 2 \delta v_{\text{dark}}(t) \cos(\omega_{0}(t-\tau)) \right] \\
            \delta \Phi_{\text{dark}}(t)&:= \text{LPF}\left[ -2 \delta v_{\text{dark}}(t) \sin(\omega_{0}(t-\tau)) \right
            ],
        \end{split}
    \end{equation}
    Projecting the photodetector voltage onto the oscillation frequency $\omega_{0}$, we have
    \begin{equation}
        \begin{split}
            \delta v_{\text{pd}}(t) =&- \frac{\pi g q_{e}\eta^{2}\bar q_{\text{in}}^{2}}{4 v_{\pi}}(J_{0}^{2}- 2 J_{1}^{2}
            ) \cos(\omega_{0}(t - \tau)) \, \delta A_{\text{mod}}(t-\tau) \\&+ \frac{\pi v_{0}g q_{e}\eta^{2}\bar q_{\text{in}}^{2}}{4
            v_{\pi}}J_{0}^{2}\sin(\omega_{0}(t - \tau)) \, \delta \varphi_{\text{mod}}(t - \tau) \\&+ \frac{g q_{e}\eta^{2}\bar
            q_{\text{in}}}{2}\cos(\omega_{0}(t-\tau)) \delta \hat{A}_{\text{qn}}(t) \\&- \frac{g q_{e}\eta^{2}\bar q_{\text{in}}}{2}
            \sin(\omega_{0}(t-\tau)) \delta \hat{\Phi}_{\text{qn}}(t) \\&+ \cos(\omega_{0}(t-\tau)) \delta A_{\text{dark}}
            (t) \\&- \sin(\omega_{0}(t-\tau)) \delta \Phi_{\text{dark}}(t).
        \end{split}
    \end{equation}
    Writing the carrier-frequency component of the photodetector voltage as
    \begin{equation}
        v_{\text{pd}}(t) = -[\bar v_{\text{pd}}+ \delta A_{\text{pd}}(t)] \cos(\omega_{0}(t-\tau)) + \bar v_{\text{pd}}\delta
        \varphi_{\text{pd}}(t) \sin(\omega_{0}(t-\tau)),
    \end{equation}
    we identify the dimensionless phase fluctuation of the band-limited photodetector voltage to be
    \begin{equation}
        \label{eq:Ppd-full}
        \begin{split}
            \delta \varphi_{\text{pd}}[\Omega] = \frac{1}{\bar v_{\text{pd}}}\bigg( \frac{\pi v_{0}g q_{e}\eta^{2}\bar q_{\text{in}}^{2}J_{0}^{2}}{4
            v_{\pi}}e^{i\Omega \tau}\delta \varphi_{\text{mod}}[\Omega] - \frac{g q_{e}\eta^{2}\bar q_{\text{in}}}{2}\delta
            \hat{\Phi}_{\text{qn}}[\Omega] - \delta \Phi_{\text{dark}}[\Omega] \bigg).
        \end{split}
    \end{equation}

    From \cref{eq:filter}, the filter acts on the full sideband frequency $\omega_{0}+\Omega$, while the photodetector quadratures
    above are written in the delayed carrier basis $\omega_{0}(t-\tau)$. Converting back to the modulator basis
    therefore contributes a factor of $e^{i\omega_0\tau}$, so the phase fluctuations of the modulation voltage are
    \begin{equation}
        \label{eq:Pmod-loop}
        \begin{split}
            \delta \varphi_{\text{mod}}[\Omega] = \frac{\bar v_{\text{pd}}}{v_{0}}H_{\text{elec}}[\omega_{0}+ \Omega] e^{i\omega_0
            \tau}\delta \varphi_{\text{pd}}[\Omega].
        \end{split}
    \end{equation}
    Substituting \cref{eq:Ppd-full} into this expression gives
    \begin{equation}
        \begin{split}
            \delta \varphi_{\text{mod}}[\Omega] = \frac{H_{\text{elec}}[\omega_{0}+ \Omega] e^{i\omega_0 \tau}}{v_{0}}\bigg
            ( \frac{\pi v_{0}g q_{e}\eta^{2}\bar q_{\text{in}}^{2}J_{0}^{2}}{4 v_{\pi}}e^{i\Omega \tau}\delta \varphi_{\text{mod}}
            [\Omega] - \frac{g q_{e}\eta^{2}\bar q_{\text{in}}}{2}\delta \hat{\Phi}_{\text{qn}}[\Omega] - \delta \Phi_{\text{dark}}
            [\Omega] \bigg).
        \end{split}
    \end{equation}
    We can write this expression in terms of the oscillator's open loop transfer function as
    \begin{equation}
        \delta \varphi_{\text{mod}}[\Omega] = - H_{\text{OL}}[\omega_{0}+ \Omega] J_{0}^{2}\delta \varphi_{\text{mod}}[ \Omega
        ] - \frac{H_{\text{elec}}[\omega_{0}+ \Omega] e^{i\omega_0 \tau}}{v_{0}}\bigg( \frac{g q_{e}\eta^{2}\bar q_{\text{in}}}{2}
        \delta \hat{\Phi}_{\text{qn}}[\Omega] + \delta \Phi_{\text{dark}}[\Omega] \bigg).
    \end{equation}

    In this model, the modulator is modeled as the oscillator's sole nonlinear element, so
    $|H_{\text{OL}}[\omega_{0}+ \Omega]| > 1$ and saturation will cause $J_{0}^{2}$ to decrease until
    $|H_{\text{OL}}[\omega_{0}+ \Omega]|J_{0}^{2}= 1$. We denote the saturated open-loop transfer function by $H_{\text{OL}}
    [\omega_{0}+ \Omega]/|H_{\text{OL}}[\omega_{0}]|$, such that the magnitude of the saturated open-loop transfer
    function is identically unity at the oscillation frequency. We can then write the phase noise of the modulation voltage
    as
    \begin{equation}
        \label{eq:noiseCondensed}\delta \varphi_{\text{mod}}[\Omega] = -\frac{H_{\text{elec}}[\omega_{0}+ \Omega] H_{\text{CL}}[\omega_{0}+
        \Omega] e^{i\omega_0 \tau}}{v_{0}}\bigg( \frac{g q_{e}\eta^{2}\bar q_{\text{in}}}{2}\delta \hat{\Phi}_{\text{qn}}
        [\Omega] + \delta \Phi_{\text{dark}}[\Omega] \bigg).
    \end{equation}
    Here, $H_{\text{CL}}[\Omega]$ is the saturated closed-loop transfer function defined by
    \begin{equation}
        H_{\text{CL}}[\Omega] = \frac{1}{1 + \frac{H_{\text{OL}}[\Omega]}{|H_{\text{OL}}[\omega_{0}]|}}.
    \end{equation}

    \subsection{Quantum Noise}

    From \cref{eq:deltaQqnDef,eq:deltaPQqnDef}, we can determine the quantum noise contribution to the oscillator's phase
    noise in terms of the optical state's quantum fluctuations. By definition,
    \begin{equation}
        \begin{split}
            \delta \hat{\Phi}_{\text{qn}}[\Omega] = \int_{-\infty}^{\infty}dt e^{i\Omega t}\, \text{LPF}\bigg\{ -2 \bigg(
            &\bigg[ J_{0}^{2}+ 2 J_{1}^{2}- 4 J_{0}J_{1}\cos(\omega_{0}(t - \tau)) + 2 J_{1}^{2}\cos(2\omega_{0}(t - \tau
            )) \bigg] \delta \hat q_{\text{in}}(t-\tau) \\ +&\bigg[ J_{0}^{2}- 2 J_{1}^{2}- 2 J_{1}^{2}\cos(2\omega_{0}(t
            - \tau)) \bigg] \delta \hat q_{\text{eoam}}^{\text{vac}}(t-\tau) \\ +&\sqrt{\frac{2(1-\eta^{2})}{ \eta^{2}}}\bigg
            [ J_{0}- 2 J_{1}\cos(\omega_{0}(t - \tau)) \bigg] \, \delta \hat q_{\text{loss}}^{\text{vac}}(t) \bigg ) \sin
            (\omega_{0}(t-\tau)) \bigg\}.
        \end{split}
    \end{equation}
    From which we find
    \begin{equation}
        \begin{split}
            \delta \hat{\Phi}_{\text{qn}}[\Omega] =&i e^{i\Omega\tau}H_{\text{lpf}}[\Omega]\left[ (J_{0}^{2}+ J_{1}^{2}) (
            \delta \hat q_{\text{in}}[\Omega + \omega_{0}] - \delta \hat q_{\text{in}}[\Omega - \omega_{0}]) + (J_{0}^{2}
            - J_{1}^{2})(\delta \hat q_{\text{eoam}}^{\text{vac}}[\Omega + \omega_{0}] - \delta \hat q_{\text{eoam}}^{\text{vac}}
            [\Omega - \omega_{0}]) \right] \\&- 2 i J_{0}J_{1}e^{i\Omega\tau}H_{\text{lpf}}[\Omega]\left( \delta \hat{q}_{\text{in}}
            [\Omega + 2\omega_{0}] - \delta \hat{q}_{\text{in}}[\Omega - 2\omega_{0}] \right) \\&+ i J_{1}^{2}e^{i\Omega\tau}
            H_{\text{lpf}}[\Omega]\big( \delta \hat q_{\text{in}}[\Omega + 3\omega_{0}] - \delta \hat q_{\text{in}}[\Omega
            - 3\omega_{0}] - \delta \hat q_{\text{eoam}}^{\text{vac}}[\Omega + 3\omega_{0}] + \delta \hat q_{\text{eoam}}
            ^{\text{vac}}[\Omega - 3\omega_{0}] \big) \\&+ i J_{0}\sqrt{\frac{2(1-\eta^{2})}{\eta^{2}}}\, H_{\text{lpf}}[
            \Omega] \left( e^{-i\omega_0 \tau}\delta \hat q_{\text{loss}}^{\text{vac}}[\Omega + \omega_{0}] - e^{i\omega_0
            \tau}\delta \hat q_{\text{loss}}^{\text{vac}}[\Omega - \omega_{0}] \right) \\&- i J_{1}\sqrt{\frac{2(1-\eta^{2})}{\eta^{2}}}
            \, H_{\text{lpf}}[\Omega] \left( e^{- 2 i\omega_0 \tau}\delta \hat q_{\text{loss}}^{\text{vac}}[\Omega + 2\omega
            _{0}] - e^{2 i \omega_0 \tau}\delta \hat q_{\text{loss}}^{\text{vac}}[\Omega - 2\omega_{0}] \right).
        \end{split}
    \end{equation}
    From this expression, we can read off the phase noise PSD due to quantum noise, which is
    \begin{equation}
        \begin{split}
            \bar{S}_{\Phi \Phi}^{\text{qn}}[\Omega] =&|H_{\text{lpf}}[\Omega]|^{2}\left[ (J_{0}^{2}+ J_{1}^{2})^{2}(\bar{S}
            _{qq}^{\text{in}}[\Omega + \omega_{0}] + \bar{S}_{qq}^{\text{in}}[\Omega - \omega_{0}]) + (J_{0}^{2}- J_{1}^{2}
            )^{2}(\bar{S}_{qq}^{\text{eoam}}[\Omega + \omega_{0}] + \bar{S}_{qq}^{\text{eoam}}[\Omega - \omega_{0}]) \right
            ] \\&+ 4 (J_{0}J_{1})^{2}|H_{\text{lpf}}[\Omega]|^{2}\left( \bar{S}_{qq}^{\text{in}}[\Omega + 2\omega_{0}] + \bar
            {S}_{qq}^{\text{in}}[\Omega - 2\omega_{0}] \right) \\&+ J_{1}^{4}|H_{\text{lpf}}[\Omega]|^{2}\big( \bar{S}_{qq}
            ^{\text{in}}[\Omega + 3\omega_{0}] + \bar{S}_{qq}^{\text{in}}[\Omega - 3\omega_{0}] + \bar{S}_{qq}^{\text{eoam}}
            [\Omega + 3\omega_{0}] + \bar{S}_{qq}^{\text{eoam}}[\Omega - 3\omega_{0}] \big) \\&+ J_{0}^{2}\left(\frac{2(1-\eta^{2})}{\eta^{2}}
            \right) \, |H_{\text{lpf}}[\Omega]|^{2}\left( \bar{S}_{qq}^{\text{loss}}[\Omega + \omega_{0}] + \bar{S}_{qq}^{\text{loss}}
            [\Omega - \omega_{0}] \right) \\&+ J_{1}^{2}\left( \frac{2(1-\eta^{2})}{\eta^{2}}\right) \, |H_{\text{lpf}}[\Omega
            ]|^{2}\left( \bar{S}_{qq}^{\text{loss}}[\Omega + 2\omega_{0}] + \bar{S}_{qq}^{\text{loss}}[\Omega - 2\omega_{0}
            ] \right).
        \end{split}
    \end{equation}
    We are concerned with phase fluctuations of the RF carrier such that $\Omega \ll \omega_{0}$, so we will drop the
    $|H_{\text{lpf}}[\Omega]|^{2}$ terms from now on and leave these implicit. Furthermore, we assume that the optical
    quadrature fluctuation level does not depend strongly on $\Omega$ for $\Omega \ll \omega_{0}$ and that these are
    symmetric with respect to the optical carrier such that
    $\bar{S}_{qq}[\Omega - n\omega_{0}] \approx \bar{S}_{qq}[\Omega + n\omega_{0}] \approx \bar{S}_{qq}[n\omega_{0}]$ (here
    $n \in \{1,2,3\}$). The above expression then simplifies to
    \begin{equation}
        \begin{split}
            \bar{S}_{\Phi \Phi}^{\text{qn}}[\Omega] =&2(J_{0}^{2}+ J_{1}^{2})^{2}\bar{S}_{qq}^{\text{in}}[\omega_{0}] + 2
            (J_{0}^{2}- J_{1}^{2})^{2}\bar{S}_{qq}^{\text{eoam}}[\omega_{0}] \\&+ 8 (J_{0}J_{1})^{2}\bar{S}_{qq}^{\text{in}}
            [2\omega_{0}] \\&+ 2 J_{1}^{4}\big( \bar{S}_{qq}^{\text{in}}[3\omega_{0}] + \bar{S}_{qq}^{\text{eoam}}[3\omega
            _{0}]\big) \\&+ 4 J_{0}^{2}\left(\frac{1-\eta^{2}}{\eta^{2}}\right) \bar{S}_{qq}^{\text{loss}}[\omega_{0}] \\
            &+ 4 J_{1}^{2}\left( \frac{1-\eta^{2}}{\eta^{2}}\right) \bar{S}_{qq}^{\text{loss}}[2\omega_{0}] .
        \end{split}
    \end{equation}

    In our experiment, we inject squeezing through the loss port such that $\bar{S}_{qq}^{\text{i}}[n \omega_{0}] = 1/2$ for
    $i \in \{ \text{in}, \text{eoam}\}$ and $n \in \{1,2,3\}$. We have
    \begin{equation}
        \label{eq:quantumNoise}\bar{S}_{\Phi \Phi}^{\text{qn}}[\Omega] = 2\left( J_{0}^{4}+ 2 J_{0}^{2}J_{1}^{2}+ 2 J_{1}
        ^{4}\right) + 4 \left(\frac{1-\eta^{2}}{\eta^{2}}\right) \left( J_{0}^{2}\bar{S}_{qq}^{\text{loss}}[\omega_{0}] +
        J_{1}^{2}\bar{S}_{qq}^{\text{loss}}[2\omega_{0}] \right).
    \end{equation}

    \subsection{Dark Noise}

    Here, we briefly compute the RF tone's phase fluctuations due to dark noise in terms of the PSD of the electronics'
    voltage noise. We find
    \begin{equation}
        \delta \Phi_{\text{dark}}[\Omega] = i H_{\text{lpf}}[\Omega]\left( e^{-i\omega_0 \tau}\delta v_{\text{dark}}[\Omega
        + \omega_{0}] - e^{i\omega_0 \tau}\delta v_{\text{dark}}[\Omega - \omega_{0}] \right),
    \end{equation}
    from which we have
    \begin{equation}
        \bar S_{\Phi \Phi}^{\text{dark}}[\Omega] = |H_{\text{lpf}}[\Omega]|^{2}\left(\bar{S}_{VV}^{\text{dark}}[\Omega + \omega
        _{0}] + \bar{S}_{VV}^{\text{dark}}[\Omega - \omega_{0}] \right).
    \end{equation}
    As in the quantum noise section, we will leave the low-pass filer implicit and we will assume that the dark noise
    voltage spectrum is symmetric about zero frequency. We have:
    \begin{equation}
        \bar S_{\Phi \Phi}^{\text{dark}}[\Omega] = \bar{S}_{VV}^{\text{dark}}[\omega_{0}+ \Omega] + \bar{S}_{VV}^{\text{dark}}
        [\omega_{0}- \Omega].
    \end{equation}

    \section{Full Analytic Noise Model}

    From \cref{eq:noiseCondensed,eq:quantumNoise}, the OEO's phase quadrature spectrum is given by
    \begin{equation}
        \begin{split}
            \bar{S}_{\varphi \varphi}^{\text{mod}}[\Omega] =&\frac{|H_{\text{elec}}[\omega_{0}+ \Omega]|^{2}|H_{\text{CL}}[\omega_{0}+
            \Omega]|^{2}}{v_{0}^{2}}\\&\times \Bigg[ \frac{g^{2}q_{e}^{2}\eta^{2}P_{\text{in}}}{\hbar \omega_{\ell}}\bigg
            ( \eta^{2}\left( J_{0}^{4}+ 2 J_{0}^{2}J_{1}^{2}+ 2 J_{1}^{4}\right) + 2 \left( 1-\eta^{2}\right) \left ( J_{0}
            ^{2}\bar{S}_{qq}^{\text{loss}}[\omega_{0}] + J_{1}^{2}\bar{S}_{qq}^{\text{loss}}[2\omega_{0}] \right ) \bigg)
            \\&\qquad + \bar{S}_{VV}^{\text{dark}}[\omega_{0}+ \Omega] + \bar{S}_{VV}^{\text{dark}}[\omega_{0}- \Omega] \Bigg
            ].
        \end{split}
    \end{equation}
    We inject squeezing through the ``loss'' port. To account for losses on this path, we let $\bar{S}_{qq}^{\text{loss}}
    = \eta_{\text{sqz}}^{2}\bar{S}_{qq}^{\text{sqz}}+ (1-\eta_{\text{sqz}}^{2})\bar{S}_{qq}^{\text{vac}}= \eta_{\text{sqz}}
    ^{2}\bar{S}_{qq}^{\text{sqz}}+ (1-\eta_{\text{sqz}}^{2})/2$. With this definition,
    \begin{equation}
        \begin{split}
            \bar{S}_{\varphi \varphi}^{\text{mod}}[\Omega] =&\frac{|H_{\text{elec}}[\omega_{0}+ \Omega]|^{2}|H_{\text{CL}}[\omega_{0}+
            \Omega]|^{2}}{v_{0}^{2}}\\&\times \Bigg[ \frac{g^{2}q_{e}^{2}\eta^{2}P_{\text{in}}}{\hbar \omega_{\ell}}\bigg
            ( \eta^{2}\left( J_{0}^{4}+ 2 J_{0}^{2}J_{1}^{2}+ 2 J_{1}^{4}\right) + \left( 1-\eta^{2}- \eta_{\text{sqz}}^{2}
            \right) \left( J_{0}^{2}+ J_{1}^{2}\right) \\&\qquad + 2 \eta_{\text{sqz}}^{2}\left( J_{0}^{2}\bar{S}_{qq}^{\text{sqz}}
            [\omega_{0}] + J_{1}^{2}\bar{S}_{qq}^{\text{sqz}}[2\omega_{0}] \right) \bigg) \\&\qquad + \bar{S}_{VV}^{\text{dark}}
            [\omega_{0}+ \Omega] + \bar{S}_{VV}^{\text{dark}}[\omega_{0}- \Omega] \Bigg].
        \end{split}
    \end{equation}
    Experimentally, $\eta^{2}$ is the transmission efficiency (in power units) from the output of the EOAM to the
    photodetector in the absence of modulation (including imperfect quantum efficiency) and $\eta_{\text{sqz}}^{2}$ is
    the transmission efficiency (in power units) from the squeezed light source to the photodetector (including imperfect
    quantum efficiency but excluding dark noise clearance).

    Since the oscillator's output voltage is proportional to its modulation voltage, we have $\bar{S}_{\varphi \varphi}^{\text{out}}
    [\Omega] = (v_{0}/v_{\text{out}})^{2}\bar{S}_{\varphi \varphi}^{\text{mod}}[\Omega]$. The oscillator's output phase
    noise is then
    \begin{equation}
        \begin{split}
            \bar{S}_{\varphi \varphi}^{\text{out}}[\Omega] =&\frac{|H_{\text{elec}}[\omega_{0}+ \Omega]|^{2}|H_{\text{CL}}[\omega_{0}+
            \Omega]|^{2}}{v_{\text{out}}^{2}}\\&\times \Bigg[ \frac{g^{2}q_{e}^{2}\eta_{\text{qe}}^{2}P_{0}}{\hbar
            \omega_{\ell}}\bigg( \eta^{2}\left( J_{0}^{4}+ 2 J_{0}^{2}J_{1}^{2}+ 2 J_{1}^{4}\right) + \left( 1-\eta^{2}- \eta
            _{\text{sqz}}^{2}\right) \left( J_{0}^{2}+ J_{1}^{2}\right) \\&\qquad + 2 \eta_{\text{sqz}}^{2}\left ( J_{0}^{2}
            \bar{S}_{qq}^{\text{sqz}}[\omega_{0}] + J_{1}^{2}\bar{S}_{qq}^{\text{sqz}}[2\omega_{0}] \right) \bigg ) \\&\qquad
            + \bar{S}_{VV}^{\text{dark}}[\omega_{0}+ \Omega] + \bar{S}_{VV}^{\text{dark}}[\omega_{0}- \Omega ] \Bigg],
        \end{split}
    \end{equation}
    where $P_{0}:= (\eta^{2}/\eta^{2}_{\text{qe}}) P_{\text{in}}$ is the power on the photodetector in the absence of
    modulation\footnote{Note that $\eta^{2}= \eta^{2}_{\text{qe}}\eta_{\ell}^{2}$ includes the effects of imperfect
    quantum efficiency so the measured optical power is $P_{0}:= (\eta^{2}/\eta^{2}_{\text{qe}}) P_{\text{in}}$, not $P_{0}
    := \eta^{2}P_{\text{in}}$. For a photodetector with unit quantum efficiency, these coincide.}.

    In our experiment, we have $J_{1}\ll 1$ and $\eta^{2}\ll 1$, so we can neglect terms of order $J_{1}^{4}$ and $\eta^{2}
    J_{1}^{2}$. We then have
    \begin{equation}
        \label{eq:fullNoiseModel}
        \begin{split}
            \bar{S}_{\varphi \varphi}^{\text{out}}[\Omega] =&\frac{|H_{\text{elec}}[\omega_{0}+ \Omega]|^{2}|H_{\text{CL}}[\omega_{0}+
            \Omega]|^{2}}{v_{\text{out}}^{2}}\\&\times \Bigg[ \frac{g^{2}q_{e}^{2}\eta_{\text{qe}}^{2}P_{0}}{\hbar
            \omega_{\ell}}\bigg( \left(1 - \eta_{\text{sqz}}^{2}\right) \left( J_{0}^{2}+ J_{1}^{2}\right) + 2 \eta_{\text{sqz}}
            ^{2}\left( J_{0}^{2}\bar{S}_{qq}^{\text{sqz}}[\omega_{0}] + J_{1}^{2}\bar{S}_{qq}^{\text{sqz}}[2\omega_{0}] \right
            ) \bigg) \\&\qquad + \bar{S}_{VV}^{\text{dark}}[\omega_{0}+ \Omega] + \bar{S}_{VV}^{\text{dark}}[\omega_{0}- \Omega
            ] \Bigg].
        \end{split}
    \end{equation}
    When there is no squeezed light injected, the field on the ``sqz'' port is vacuum and $\bar{S}_{qq}^{\text{sqz}}[\omega
    _{0}] = \bar{S}_{qq}^{\text{sqz}}[2 \omega_{0}] = 1/2$. In this case, the output phase noise further simplifies to
    \begin{equation}
        \label{eq:fullNoiseModelShot}
        \begin{split}
            \bar{S}_{\varphi \varphi}^{\text{out}}[\Omega] =&\frac{|H_{\text{elec}}[\omega_{0}+ \Omega]|^{2}|H_{\text{CL}}[\omega_{0}+
            \Omega]|^{2}}{v_{\text{out}}^{2}}\Bigg( \frac{g^{2}q_{e}^{2}\eta_{\text{qe}}^{2}P_{0}}{\hbar \omega_{\ell}}\left
            ( J_{0}^{2}+ J_{1}^{2}\right) + \bar{S}_{VV}^{\text{dark}}[\omega_{0}+ \Omega] + \bar{S}_{VV}^{\text{dark}}[\omega
            _{0}- \Omega] \Bigg).
        \end{split}
    \end{equation}
    Experimentally, we find that $|H_{\text{elec}}[\omega_{0}+ \Omega]|$ is approximately constant within 100 kHz of the oscillator's resonant frequency with a value of $|H_{\text{elec}}[\omega_{0}+ \Omega]| \approx 0.62$. We use this value when comparing our noise model to the OEO's measured phase noise.

    \section{A standard quantum limit for opto-electronic oscillators}

    The standard quantum limit (SQL) of an OEO can be evaluated readily using our existing analysis of the OEO's mean-field
    and noise response. Here, we define the SQL of an OEO to be the lower bound on the free-running phase noise of the
    oscillator set by vacuum fluctuations in the optical carrier supporting the loop. To see how the SQL arises, we
    combine the results in \cref{eq:expOscCond,eq:fullNoiseModel} by substituting the analytical expression for the RF
    mean-field, $v_{0}$, into the noise model. From this we obtain a modified version of \cref{eq:fullNoiseModel} as follows:

    \begin{equation}
        \label{eq:sqlTerms1}
        \begin{split}
            \bar{S}_{\varphi \varphi}^{\text{out}}[\Omega] =&|H_{\text{CL}}[\omega_{0}+ \Omega]|^{2}\\&\times \Bigg[ \frac{\hbar
            \omega_{\ell}}{\eta_{qe}P_{0}J_{1}(2\beta)^{2}}\bigg( \left( 1-\eta_{\text{sqz}}^{2}\right) \left( J_{0}^{2}+
            J_{1}^{2}\right) \\&\qquad + 2 \eta_{\text{sqz}}^{2}\left( J_{0}^{2}\bar{S}_{qq}^{\text{sqz}}[\omega_{0}] + J
            _{1}^{2}\bar{S}_{qq}^{\text{sqz}}[2\omega_{0}] \right) \bigg) \\&\qquad + \frac{\hbar^{2}\omega_{\ell}^{2}}{q_{e}^{2}P_{0}^{2}g^{2}}
            \left(\bar{S}_{VV}^{\text{dark}}[\omega_{0}+ \Omega] + \bar{S}_{VV}^{\text{dark}}[\omega_{0}- \Omega]\right) \Bigg
            ].
        \end{split}
    \end{equation}

    \noindent
    Here, $g$ is explicitly the unsaturated transimpedance amplifier gain, However, in the case where both the modulator
    and the transimpedance amplifier saturate, $g$ can be implicitly understood as the saturated gain of the transimpedance amplifier.
    Furthermore, if we assume that the loss port is driven by vacuum fluctuations such that $\bar{S}_{qq}^{\text{sqz}}[\omega
    _{0}] = \bar{S}_{qq}^{\text{sqz}}[2\omega_{0}] = 1/2$ and that the dark noise is symmetric about $\omega_{0}$ such that
    $\bar{S}_{VV}^{\text{dark}}[\omega_{0}+ \Omega] = \bar{S}_{VV}^{\text{dark}}[\omega_{0}- \Omega]$, then the OEO's
    phase noise can be expressed in the standard case as

    \begin{equation}
        \label{eq:sqlTerms2}\bar{S}_{\varphi \varphi}^{\text{out}}[\Omega] = |H_{\text{CL}}[\omega_{0}+ \Omega]|^{2}
        \Bigg[ \frac{\hbar \omega_{\ell}}{\eta_{qe}P_{0}}\bigg(\frac{J_{0}(\beta)^{2}+ J_{1}(\beta)^{2}}{J_{1}(2\beta)^{2}}
        \Bigg) + \frac{2\hbar^{2}\omega_{\ell}^{2}}{q_{e}^{2}P_{0}^{2}g^{2}J_{1}(2\beta)^{2}}\bar{S}_{VV}^{\text{dark}}[\omega
        _{0}+ \Omega] \Bigg].
    \end{equation}

    \noindent
    In \cref{eq:sqlTerms2} the phase noise is driven by two terms that scale differently with the optical carrier power
    on the in-loop photodetector. The terms scaling with $1/P_{0}$ are physically understood to be the shot-noise scaling
    owed to the optical vacuum fluctuations present on the detector. Terms scaling with $1/P_{0}^{2}$ correspond to the
    effect of dark noise on the oscillator's phase noise. From this, it is apparent that with sufficient optical power, any
    OEO will become limited by vacuum fluctuations as the dark noise contributions to the phase noise decay.

    Now, we define the standard-quantum-limited phase noise of the oscillator neglecting dark-noise contributions and assuming unit quantum detection efficiency ($\eta_{qe}$=1) as $\bar{S}_{\varphi \varphi}^{\text{SQL}}[\Omega]$, such that

    \begin{equation}\label{eq:sqlFinal}
        \bar{S}_{\varphi \varphi}^{\text{out}}[\Omega] \ > \bar{S}_{\varphi \varphi}^{\text{SQL}}[\Omega
        ] = |H_{\text{CL}}[\omega_{0}+ \Omega]|^{2}\frac{\hbar \omega_{\ell}}{P_{0}}\bigg(\frac{J_{0}(\beta)^{2}+ J_{1}(\beta)^{2}}{J_{1}(2\beta)^{2}}
        \bigg) \approx \frac{\hbar \omega_{\ell}}{P_{0}\Omega^{2}\tau_{g}^{2}}\bigg(\frac{J_{0}(\beta)^{2}+ J_{1}(\beta)^{2}}{J_{1}(2\beta)^{2}}
        \bigg)
    \end{equation}

    \noindent
    using the approximation
    $|H_{\text{CL}}[\omega_{0}+ \Omega]|^{2}\approx 1/(\Omega^{2}\tau_{g}^{2})$. $\tau_{g}$ is the full group delay
    of the loop including any electronic filtering of the band-limited photocurrent. Here, we define $\tau_{g}= -d\Phi(H_{OL}
    [\omega])/d\omega$ at the oscillator's steady-state output frequency $\omega_{0}$. Finally, in the small-saturation regime
    near the oscillator's threshold, the Bessel functions in \cref{eq:sqlFinal} can be approximated such that
    $J_{0}(\beta) \approx 1$ and $J_{1}(\beta) \approx \beta/2$ to yield a more digestible form of the SQL that exposes
    the most relevant system parameters

    \begin{equation}\label{eq:sqlApprox}
        \bar{S}_{\varphi \varphi}^{\text{SQL}}[\Omega] = \frac{\hbar \omega_{\ell}}{P_{0}\Omega^{2}\tau_{g}^{2}}\bigg(
        \frac{1}{\beta^{2}} + \frac{1}{2} \bigg) = \frac{\hbar \omega_{\ell}}{P_{0}\Omega^{2}\tau_{g}^{2}}\bigg(\frac{4
        v_{\pi}^{2}}{v_{0}^{2}} + \frac{1}{2} \bigg) .
    \end{equation}

    \noindent
    Like the noise model, this SQL expression is only valid in the limit where the modulator is weakly compressing ($\beta
    < 1$). Nevertheless, we find a Schawlow-Townes-like phase noise limit \cite{SchaTow58} which improves with increased optical
    power and RF output power. We see that the expressions in \cref{eq:sqlFinal,eq:sqlApprox} contain the sum of two terms. The first of these, $(J_0(\beta)/J_1(2\beta))^2$ in \cref{eq:sqlFinal} and $1/\beta^2$ in \cref{eq:sqlApprox}, arise from the optical carrier's shot noise. The latter, $(J_1(\beta)/J_1(2\beta))^2$ in \cref{eq:sqlFinal} and $1/2$ in \cref{eq:sqlApprox}, arise from the optical sidebands' shot noise. In OEOs with very low modulation depths such that $\beta \ll 1$, the shot noise from the optical carrier completely dominates. As the modulation depth increases, the sidebands grow larger with respect to the carrier and the contribution from the sidebands' shot noise becomes increasingly important.
    This result is similar to that found in early OEO models such as in
    \cite{YaoMal96b, Romisch2000}. Interestingly, the SQL can be expressed entirely in terms of fundamental physical
    constants and the modulator's figure of merit ($v_{\pi}$).

    \section{Experimental Implementation}

\begin{figure}[!t]
        \centering
        \includegraphics[width=0.95\columnwidth]{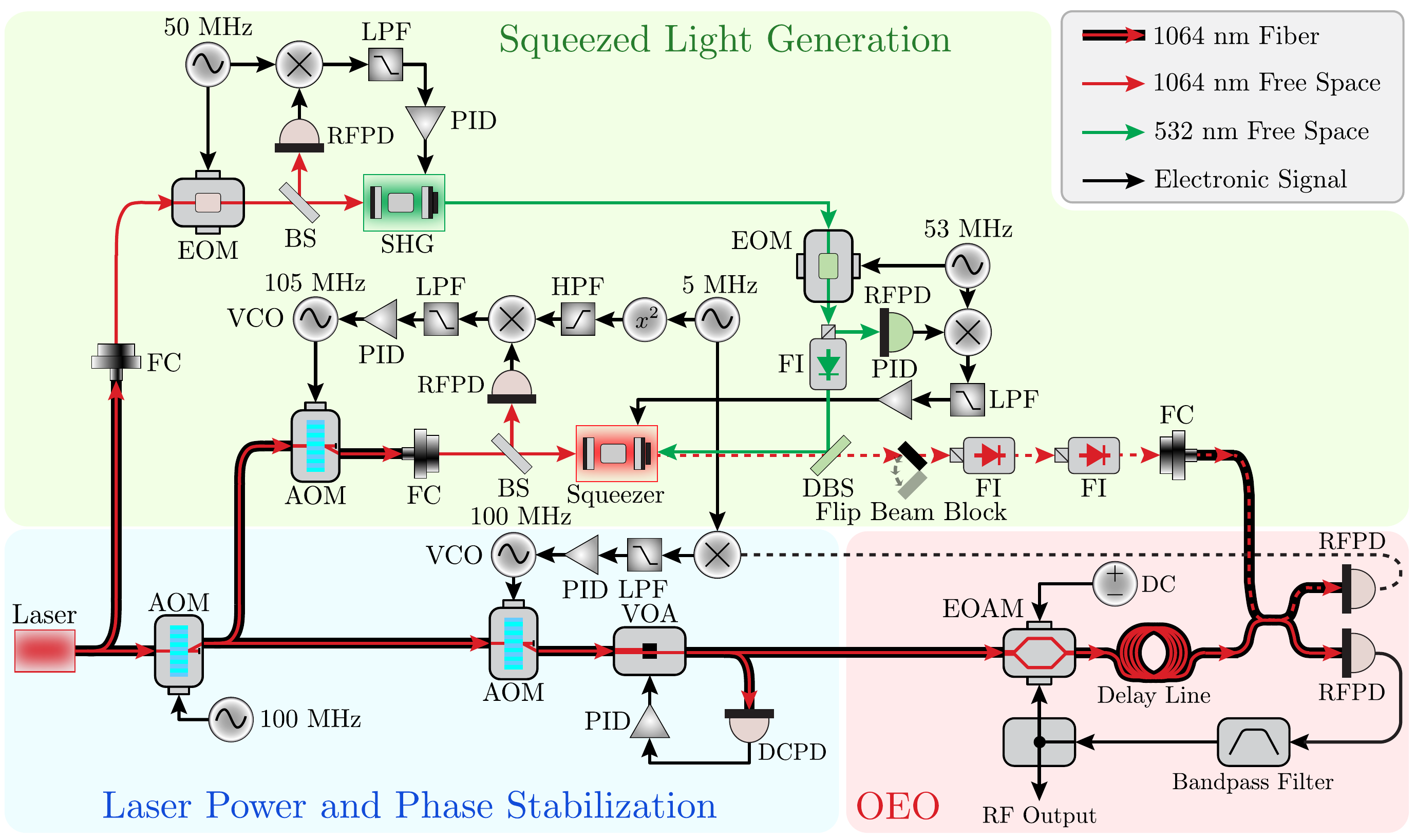}
        \caption{A schematic of the full optics and electronics paths in the experiment. As in \cref{fig:simple_schematic},
        dashed lines denote signal pathways that are only present when squeezed light is injected into the OEO. Squeezed light
        is prepared in the Squeezed Light Generation block, the laser has its phase synchronized to the squeezed beam's
        and its power stabilized in the Laser Power and Phase Stabilization block. The optical outputs of each of these blocks
        are sent to the OEO. In this schematic, FC is a fiber to free space coupler, EOM is an electro-optic phase
        modulator, BS is a beam splitter, RFPD is a radio-frequency photodetector, DCPD is a DC photodetector, LPF is a
        low pass filter, HPF is a high pass filter, PID is a proportional-integral-differential controller, SHG is a second
        harmonic generation cavity, VCO is a voltage controlled oscillator, FI is a Faraday isolator, DBS is a dichroic
        beam splitter, AOM is an acousto-optic modulator, VOA is a variable optical attenuator, DC is a constant (direct
        current) voltage source, and EOAM is an electro-optic amplitude modulator. }
        \label{fig:fullExperimentSchematic}
    \end{figure}

    In this section, we describe the details of the OEO's design and measurement. A detailed diagram of our experimental
    implementation is depicted in \cref{fig:fullExperimentSchematic}. The OEO experiment is comprised of two main subsystems,
    the OEO itself, and a squeezed-light source used to produce the squeezed state ultimately injected into the OEO system.
    Both the OEO and squeezed-light source derive their pump light from the same 1064-nm NPRO laser source to simplify
    the phase stabilization needed for squeezed-light injection.

    Our OEO is constructed similarly to the one shown in \cite{YaoMal96b}. In our implementation, the local oscillator (LO)
    power is pre-stabilized by an intensity stabilization servo to suppress its low-frequency intensity fluctuations.
    The resulting LO light is routed through the OEO's electro-optic amplitude modulator (EOAM). The EOAM's bias is held near
    the quadrature point with a constant voltage source provided to its DC-bias port terminals. The modulated light from
    the EOAM then propagates through a 30-meter-long fiber delay line and terminates into a 90\%/10\% fiber beam
    splitter that provides a port through which we inject squeezed light. This fiber beam splitter's output ports directly
    feed two fiber-coupled Newport 1811 RF photodetectors - one of these photodetectors serves as the in-loop photodetector
    for the OEO and the other serves as an out-of-loop photodetector to control the relative phase angle between the squeezed field and the LO field.
    The in-loop
    photodetector is fed from the beam-splitter port that nominally has a 90-\% transmissivity to the loss port.

    The in-loop OEO electronics consist of a band-pass loop filter (MiniCircuits BBP-10.7+) and a 3-dB RF coupler to allow
    for an RF dither to be applied to the EOAM to determine the bias voltage. The electronic gain in the OEO loop is provided entirely by
    the internal transimpedance amplifier (TIA) circuitry of the in-loop photodetector. The output of the 3-dB coupler directly
    drives the EOAM's RF port. Since our particular EOAM presents a high-impedance load to the output of the coupler, we directly
    tap the RF output of coupler at the modulator using a simple SMA tee connector and route the RF signal from the OEO to
    an Agilent N9030A PXA Signal Analyzer through 50-$\Omega$ coaxial cable for direct phase noise measurement. In this
    way, the signal analyzer serves both as instrumentation and the OEO's 50-$\Omega$ RF load. Extra care was taken to
    keep transmission line lengths in the electronic portion of the OEO loop as short as possible to allow for the RF
    components in the loop to be treated as lumped elements.

    The squeezed-light source in our experiment is a small-footprint custom source based on a linear-cavity doubly-resonant
    optical parametric oscillator (OPO). The source generates single-mode (degenerate) squeezed-vacuum light \cite{galiana_development_2022}.
    In this system (illustrated by \cref{fig:fullExperimentSchematic}), the 1064-nm pump laser drives a second-harmonic-generation
    (SHG) cavity that serves as a 532-nm pump source for the OPO. Using standard coherent control techniques
    \cite{Vahlbruch06}, a 1064-nm coherent locking field (CLF) detuned by 5 MHz from the pump laser is injected into the OPO through its high-reflectivity mirror and is phase-locked to the squeezing
    phase angle using an RF phase-locked loop (PLL). CLF phase actuation is achieved by voltage-controlled tuning of the
    CLF's up-shift acousto-optic modulator (AOM) drive frequency. The transmitted CLF, along with the generated squeezed
    light exits the output coupler of the OPO and propagates through two Faraday isolators to limit backscatter effects, ultimately
    coupling into a PM fiber that transmits the squeezed light into the loss-port of the OEO loop. The CLF sidebands
    from the OPO beat against the OEO's LO on the out-of-loop photodetector. This beat tone is locked (by actuating the
    LO up-shift AOM drive frequency) to a 5-MHz reference oscillator with another PLL to control and stabilize the squeezing
    phase angle in the OEO loop.

    \section{Setup Calibration}

    \begin{figure}[!t]
        \centering
        \includegraphics[width=0.95\columnwidth]{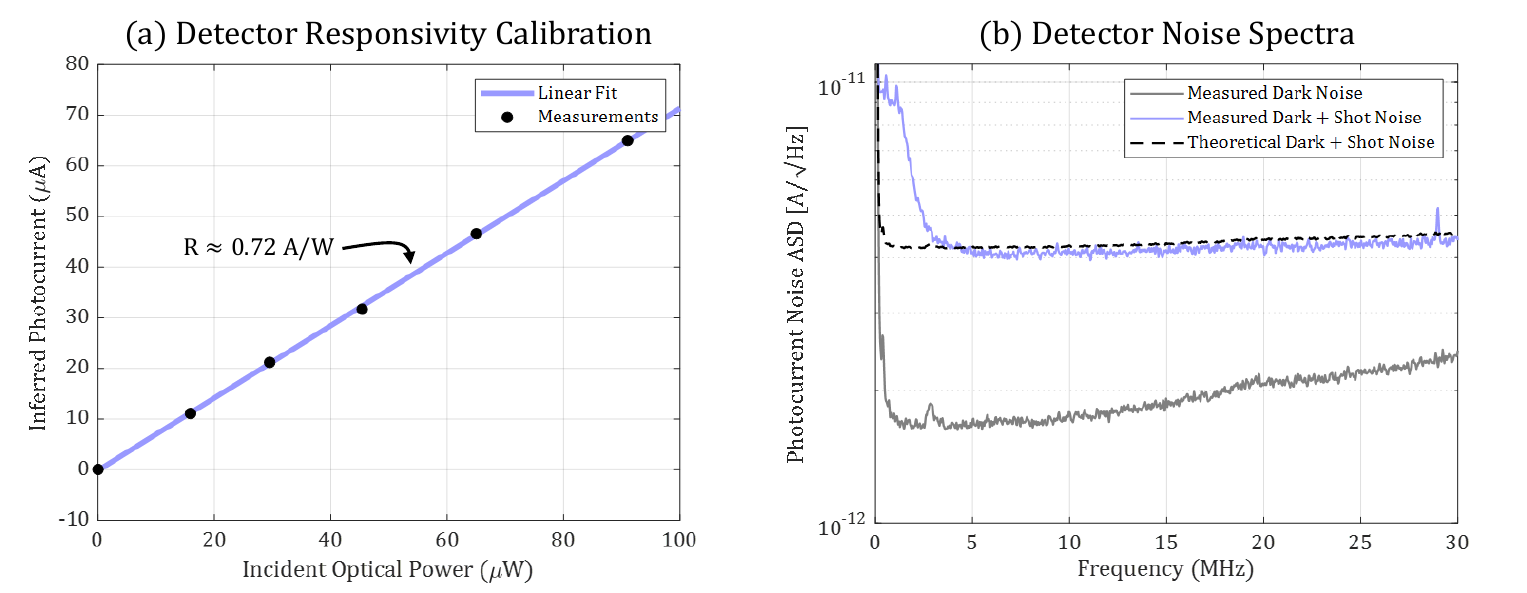}
        \caption{Characterization of the in-loop RF photodetector. (a) Power calibration of the in-loop photodetector's responsivity
        with linear fit corresponding to a responsivity of 0.72 A/W. (b) measured dark noise and vacuum shot-noise
        spectra from the in-loop photodetector with approximately 65 $\mu$W of optical power incident on the detector. }
        \label{fig:pdCharacterization}
    \end{figure}

    In this section, we describe the relevant parameters of the experiment that are required to analytically model the
    performance of our OEO. In order to evaluate \cref{eq:fullNoiseModel}, we perform a suite of measurements to provide
    accurate calibrations of the open-loop transfer function of the OEO loop ($H_{OL}[\Omega]$), the EOAM's half-wave voltage ($v_{\pi}$) and the in-loop photodetector's
    quantum efficiency ($\eta_{qe}$) and output dark noise ($\bar{S}_{VV}^{\text{dark}}[\Omega]$).

    The open-loop transfer function of the entire OEO loop is measured by breaking the loop at the RF input port of the
    EOAM in \cref{fig:fullExperimentSchematic} and measuring the system's response to a sinusoidal stimulus. Using an RF vector-network analyzer (VNA), a
    through-response measurement is taken driving the EOAM RF port and receiving on the disconnected output of the RF
    coupler that would nominally close the oscillator loop. The highly nonlinear gain-saturation characteristics of a
    feedback oscillator warrant the measurement of the loop transfer function at various large RF stimulus powers. This is
    accomplished by stepping the RF stimulus power of the VNA and recording the transfer function at each step.
    For a given frequency near the oscillator's natural frequency (9.5 MHz), the loop phase and gain dependence on the
    RF mean-field power at the EOAM's RF port are measured and characterized in \cref{fig:classical_characterization}(a).

    Notably, the gain saturation observed diverges from that predicted by the gain compression of the EOAM alone (see
    \cref{eq:oscCondition}). This suggests that our EOAM's gain saturation
    is likely dominated by the TIA internal to the in-loop photodetector. We also observe a phase shift in the open-loop
    response as the RF drive to the EOAM is increased. This justifies the particular operating LO power used in all
    measurements of the OEO's phase noise, as an LO power of approximately 65 $\mu$W results in an RF mean-field of
    approximately 6 dBm (or 4 mW) driving the EOAM. At this power, the first-order dependence of the open-loop response's
    phase on RF mean-field power (and therefore LO power) is minimized. To obtain an accurate open-loop transfer function 
    under conditions similar to those in play during oscillation, the open-loop transfer function used in
    the analytic model of the OEO is taken with an RF drive of 6 dBm to the EOAM. 

    Next, we characterize the in-loop photodetector's responsivity to determine its quantum efficiency. The Newport 1811 photodetector
    features a DC photocurrent readout which can be measured as a function of increasing incident optical power on the photodetector.
    \Cref{fig:pdCharacterization}(a) plots the inferred photocurrent generated by the photodiode in the Newport 1811 at several
    optical powers measured against a large-detector-area handheld optical power meter. A fit of this data estimates the
    responsivity of the photodiode to be 0.72 A/W. This corresponds to a quantum efficiency of 83\% at a wavelength of 1064
    nm.

    Additionally, the noise characteristics of the photodetector are measured. Two RF spectra from the AC port of the in-loop photodetector
    are measured with the same Agilent RF signal analyzer. First, one spectra is recorded to obtain the input-referred dark-noise
    current amplitude spectral density (ASD) of the detector. Next, another spectra is recorded with approximately 65 $\mu$W
    of LO power incident on the photodetector. The inferred photocurrent ASDs from this calibration
    are plotted in \cref{fig:pdCharacterization}(b). The added shot noise from the incident laser
    light on the photodetector nearly agrees with that predicted by the measured responsivity and incident optical power on
    the photodiode. The additional noise at lower frequencies is likely caused by electronic noise in the oscillators
    driving the LO-path AOMs and the relative intensity noise (RIN) of the source laser.

    \begin{figure}[!t]
        \centering
        \includegraphics[width=0.95\columnwidth]{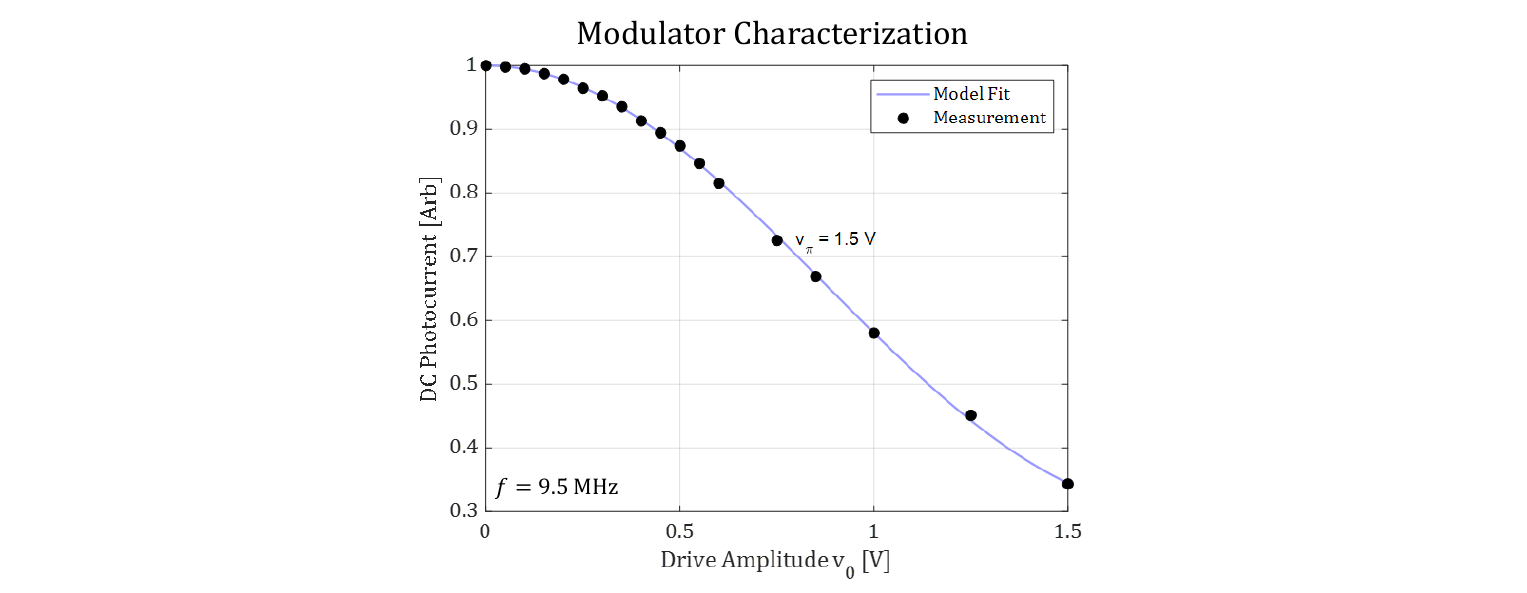}
        \caption{Characterization of the EOAM half-wave voltage using DC measurements outlined in ref. \cite{bui_method_2018}. Fit of experimental data estimates $v_{\pi} = 1.5$ V at 9.5 MHz. }
        \label{fig:modCharacterization}
    \end{figure}

    Finally, the EOAM's half-wave voltage ($v_{\pi}$) is characterized using a method similar to that outlined in \cite{bui_method_2018}. In this method,
    the modulator is driven at a frequency of 9.5 MHz and biased to peak transmission ($\phi_{bias} = 0$). The DC photocurrent drop on the in-loop photodetector is measured over a range of
    increasing drive voltages ($v_0$). At peak bias, the normalized DC photocurrent can be expressed as a function of the applied RF drive voltage \cite{bui_method_2018}

    \begin{equation}\label{eq:mod_cal}
        I^{DC}_{norm}(v_0) = \left(1 + \sqrt{1 - \delta^2}\right)^{-1}\left(1 + \sqrt{1 - \delta^2}J_0\left(
        \frac{\pi v_0}{v_{\pi}}\right)\right).
    \end{equation}

    \noindent This curve is parameterized in terms of $v_{\pi}$ and the effective splitting ratio error ($\delta$) of the Mach-Zehnder interferometer that comprises the EOAM. 
    The measured DC photocurrent is normalized and fit to \cref{eq:mod_cal}. \Cref{fig:modCharacterization} plots the measurement data and resulting model fit that estimates the EOAM's $v_{\pi}$ to be approximately 1.5 volts at a drive frequency of 9.5 MHz. The calibrated value for $v_{\pi}$ is used to evaluate the theoretical noise model for our OEO (\cref{eq:sqlTerms1}) in \cref{fig:oeoNoiseExpVsModel}.

    \section{Measurement Procedure}

    In this section, we discuss the methods and procedures used to gather consistent measurements of the OEO's phase noise
    enhancement with squeezed-light injection. Measurements of multiple experimental parameters before and during the
    phase noise measurement procedure must be recorded to determine the validity of data. These auxiliary data include initial
    tomography of the squeezed state measured at the in-loop photodetector, tracking of the vacuum phase noise in
    between measurements of squeezed phase noise, and tracking of the average LO power on the in-loop photodetector throughout
    an entire phase-noise measurement suite.

    \begin{figure}[!t]
        \centering
        \includegraphics[width=0.95\columnwidth]{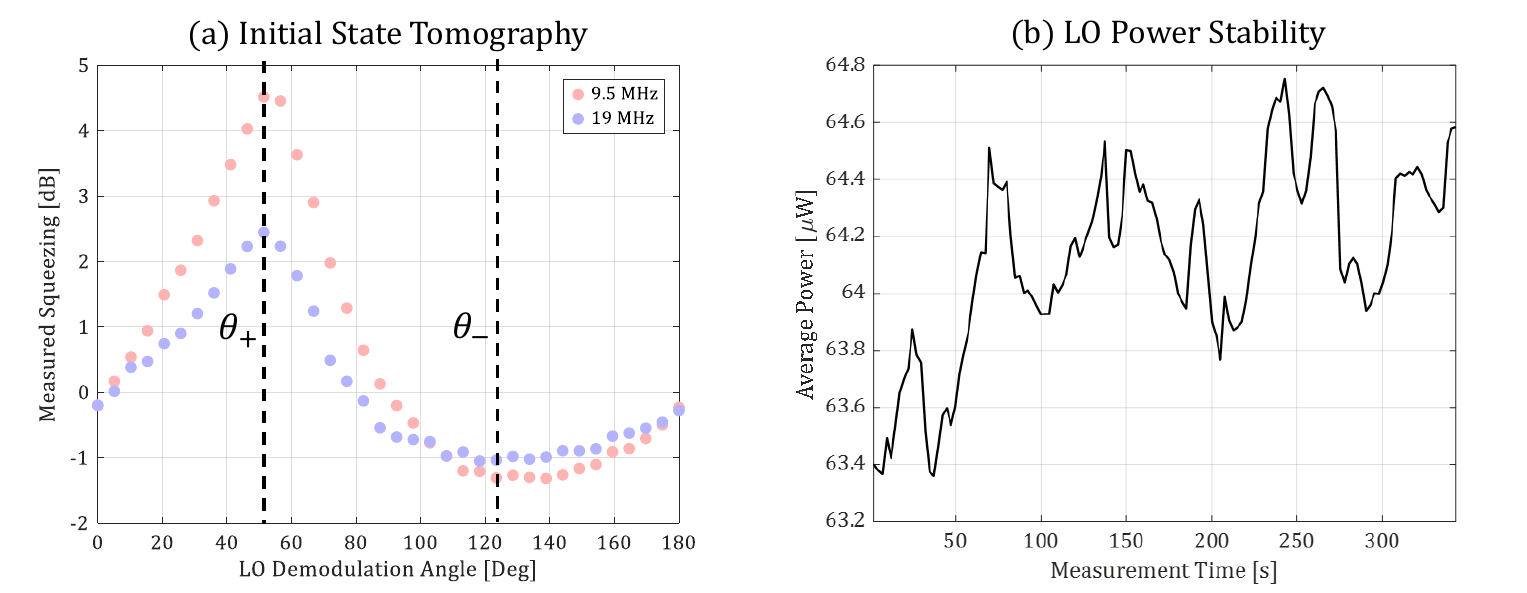}
        \caption{Characterizations of the OEO throughout the measurement procedure. (a) Initial tomography of the squeezed-state
        showing measured squeezing as a function of LO demodulation angle. (b) Measurement of the long-term LO power
        stability throughout the phase-noise measurement procedure - (a) and (b) coincide with the data in
        \cref{fig:oeoQuantumEnhancement}. }
        \label{fig:measProcedure}
    \end{figure}

    For a given group of datasets, an initial tomographic measurement of squeezing on the in-loop photodetector is made to
    estimate the OPO's generated squeezing and detection losses. This is performed once the OPO and all required phase-stabilization
    loops are locked. Additionally, the beam block shown in \cref{fig:fullExperimentSchematic} is electronically flipped
    out of the squeezed-light path. This initial tomography is performed with the oscillator loop open to stop the oscillation.
    The loop is broken at the output of the in-loop photodetector and the RF signal analyzer is connected to the in-loop
    photodetector's AC output. Next, the demodulation phase-angle of the PLL stabilizing the LO phase angle to the
    squeezing phase angle is incremented to sweep the squeezing angle on the in-loop photodetector. At each step, the
    signal analyzer acquires a 100-kHz-wide swept-spectrum measurement centered at both 9.5 MHz and 19 MHz. These constitute
    the first and second harmonic of the OEO's natural frequency. The integrated power in both bands at each
    demodulation angle is compared to a reference vacuum shot-noise power measured in each band to determine (1) the
    amount of measured squeezing and anti-squeezing on the in-loop photodetector, and (2) the LO demodulation angles corresponding
    to the maximally squeezed and anti-squeezed quadratures ($\theta_{-}$ and $\theta_{+}$ respectively). An example of
    this tomographic measurement is plotted in \cref{fig:measProcedure}(a) and corresponds to the data displayed in
    \cref{fig:oeoQuantumEnhancement}.

    From the measured minimum squeezed and maximum anti-squeezed variances relative to the vacuum photodetection variance
    ($V^{-}$ and $V^{+}$), the total detection efficiency ($\eta_{\text{sqz}}^{2}$) and generated squeezing ($\alpha^{SQZ}
    _{dB}$) at the OPO are inferred with the following two equations after neglecting the effects of LO phase noise:

    \begin{equation}
        \label{eq:tomoEff}\eta_{\text{sqz}}^{2}= \frac{(V^{+}- \gamma - 1)(1 + \gamma - V^{-})}{V^{+}+ V^{-}- 2\gamma - 2}
    \end{equation}

    \begin{equation}
        \label{eq:tomoSqz}\alpha^{SQZ}_{dB}= 10\log\left(\frac{V^{+}- \gamma - 1}{1 + \gamma - V^{-}}\right).
    \end{equation}

    \noindent
    Here, $\gamma$ is the ratio of the dark-noise PSD to the theoretically-predicted vacuum shot-noise PSD present at
    the photodetector's AC output ($\bar{S}_{VV}^{\text{dark}}/\bar{S}_{VV}^{\text{shot}}$). These two extrapolated parameters
    of the squeezed-light source and the cavity parameters of the OPO are used to inform a simple frequency-dependent model
    of the squeezing incident on our OEO loss port \cite{Dwyer13}. This model is then
    used to compute analytic values for $\bar{S}_{qq}^{\text{sqz}}[\omega_{0}+ \Omega]$ and
    $\bar{S}_{qq}^{\text{sqz}}[2\omega_{0}+ \Omega]$ in order to evaluate our analytic model of the OEO's phase noise (\cref{eq:fullNoiseModel}). From this, we find that the predicted amount of squeezing and anti-squeezing in the RF phase noise expected in our experiment agrees well with the measurement data in \cref{fig:oeoQuantumEnhancement}.

    It is assumed that the generated squeezed state is stable over a time frame of roughly one hour, so
    this tomographic extrapolation step is only required once for a group of datasets. For a more real-time measurement
    of the generated squeezing, one might consider implementing diagnostic readouts of the OPO's classical response to infer
    the generated squeezing level without requiring direct measurement of the squeezed state \cite{ganapathy_probing_2022}.

    Next, the OEO loop is closed and the RF signal analyzer is connected to the tee at the RF port of the EOAM. We experimentally
    found that the bias phase of the EOAM in the OEO would slowly thermalize due to the oscillating RF mean-field drive power
    on the device. To achieve stable LO power throughout a given dataset, we typically allow for a 20-minute warm-up period
    before the phase noise is measured. After this 20-minute window, the LO PLL is unlocked and the beam-block in \cref{fig:fullExperimentSchematic}
    is flipped to block the squeezed-light path to the OEO.

    The phase noise measurement process consists of five phase-noise measurement steps. The first measurement in this
    current state is the vacuum phase noise of the OEO. In each phase-noise measurement, the RF signal analyzer
    directly acquires a 50-trace averaged phase noise spectrum from the OEO. Each phase noise measurement takes approximately
    1 minute. After the first vacuum-noise reference spectrum is acquired, the squeezed-light beam is unblocked and the LO
    PLL is locked to the demodulation phase angle that corresponds to the maximally squeezed quadrature on the in-loop photodetector
    ($\theta_{-}$). The same phase noise measurement is then re-acquired, now with squeezed-light injection. This
    process is repeated once more but with the LO PLL locked to the phase angle that corresponds to the maximally anti-squeezed
    quadrature ($\theta_{+}$). The full measurement is concluded with one final vacuum-noise reference measurement.
    Interspersing vacuum phase noise reference spectra measurements throughout the phase noise acquisition process allows us to
    verify that the vacuum phase noise we compare the squeezed spectra to is consistent throughout the entire
    measurement.

    \begin{figure}[!t]
        \centering
        \includegraphics[width=0.95\columnwidth]{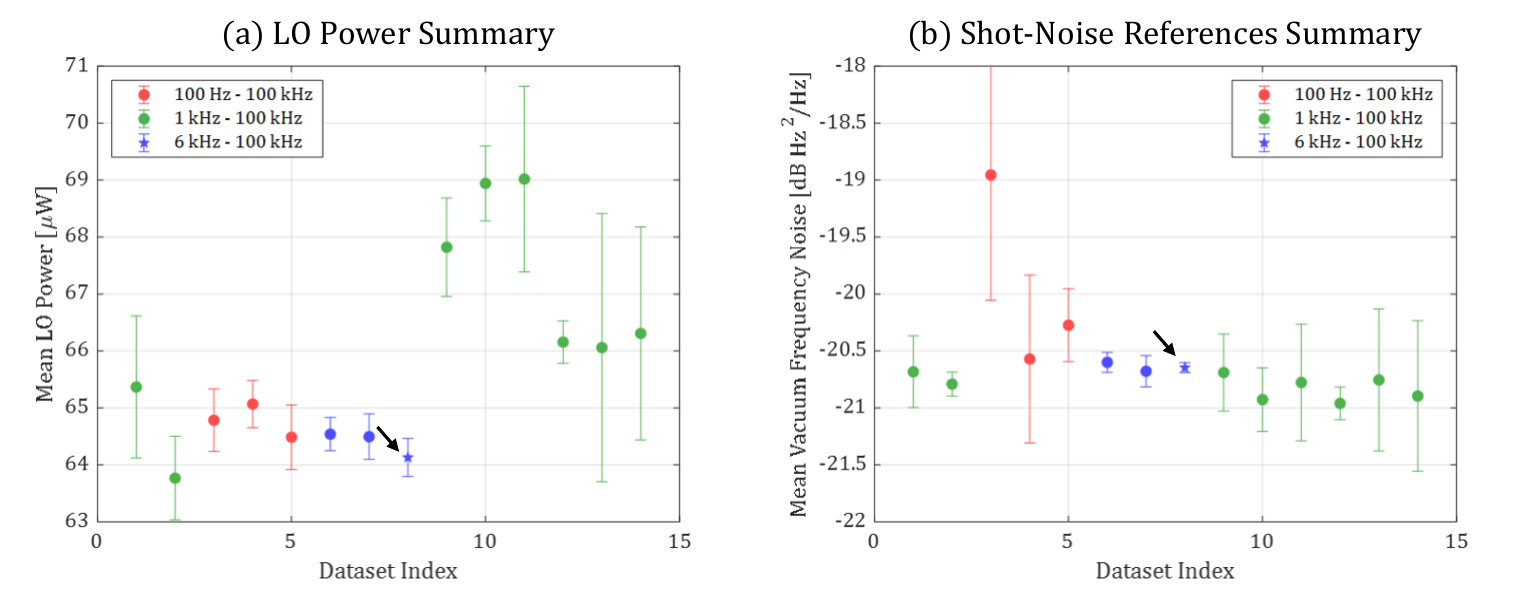}
        \caption{Plot of the mean value and standard deviation of (a) the average LO power on the in-loop photodetector
        and (b) the average reference vacuum phase noise during phase noise measurement. The x-axis indexes individual
        phase-noise measurement datasets. The mean value of each measurement is displayed with a marker; each measurement's
        standard deviation displayed with its error bar size. Blue, green, and red markers correspond to lower sideband offset frequencies of 6 kHz, 1 kHz, and 100 Hz respectively. Measurement corresponding to the data in
        \cref{fig:oeoQuantumEnhancement} highlighted with arrow.}
        \label{fig:dataSelection}
    \end{figure}

    In addition to interspersed vacuum-noise reference measurements, we use a low-speed data logger to digitize and record
    the in-loop photodetector's DC readout. With the appropriate calibration, this allows us to track the drift in
    average optical power on the in-loop photodetector. This optical power is recorded at a two-second measurement interval
    throughout the phase-noise measurement process described above. The dataset we present in \cref{fig:oeoQuantumEnhancement}
    corresponds to the data index highlighted in \cref{fig:dataSelection}. This dataset is verified to have exhibited both a
    stable average LO power on the in-loop photodetector, and a very stable vacuum phase noise reference throughout the measurement.
    The data from the remaining experiments depicted in \cref{fig:dataSelection} correspond to different phase-noise measurements
    with varying sideband offset bandwidths. Since broader bandwidth phase noise measurements require more time to complete, those datasets
    exhibit more statistical variation over the measurement than the narrow-bandwidth measurements. Because the phase noise spectrum at
    offset frequencies below 1kHz is dominated by environmental noise sources, we choose to limit the sideband offset bandwidths
    of the final measurement runs to 6 kHz in order to suppress the effects of long-term environmental drift in the high-frequency phase noise measurement.  

\end{document}